\documentclass[twocolumn, times]{aastex63}
\usepackage{multirow}
\usepackage{tabularx}

\shortauthors{Ng et al.}
\shorttitle{Galaxy--cool CGM Connections at $z<0.4$}

\usepackage{color}

\def\caii{\ion{Ca}{2}\ }
\def\mgii{\ion{Mg}{2}\ }

\begin{document}

\title{A Comprehensive Characterization of Galaxy--cool CGM Connections at $z<0.4$ \\ with DESI Year 1 Data}
\correspondingauthor{Yu Voon Ng, Ting-Wen Lan}
\email{r12244006@ntu.edu.tw, twlan@ntu.edu.tw}
\author[0009-0006-0529-5460]{Yu Voon Ng}
\affiliation{Graduate Institute of Astrophysics, National Taiwan University, No. 1, Sec. 4, Roosevelt Rd., Taipei 10617, Taiwan}

\author[0000-0001-8857-7020]{Ting-Wen Lan}
\affiliation{Graduate Institute of Astrophysics, National Taiwan University, No. 1, Sec. 4, Roosevelt Rd., Taipei 10617, Taiwan}
\affiliation{Department of Physics, National Taiwan University, No. 1, Sec. 4, Roosevelt Rd., Taipei 10617, Taiwan}
\affiliation{Institute of Astronomy and Astrophysics, Academia Sinica, No. 1, Sec. 4, Roosevelt Rd., Taipei 10617, Taiwan}

\author[0000-0002-7738-6875]{J. Xavier Prochaska}
\affiliation{Department of Astronomy and Astrophysics, University of California, Santa Cruz, 1156 High Street, Santa Cruz, CA 95065, USA}
\affiliation{Kavli Institute for the Physics and Mathematics of the Universe (Kavli IPMU), WPI, The University of Tokyo Institutes for Advanced Study (UTIAS), The
University of Tokyo, Kashiwa, Chiba, Kashiwa 277-8583, Japan}
\affiliation{Division of Science, National Astronomical Observatory of Japan, 2-21-1, Osawa, Mitaka, Tokyo 181-8588, Japan}

\author[0000-0003-4357-3450]{Am{\'e}lie Saintonge}
\affiliation{Department of Physics \& Astronomy, University College London,
Gower Street, London, WC1E 6BT, UK}

\author[0000-0002-0196-3496]{Yu-Ling Chang}
\affiliation{Graduate Institute of Astrophysics, National Taiwan University, No. 1, Sec. 4, Roosevelt Rd., Taipei 10617, Taiwan}

\author[0000-0002-2949-2155]{Małgorzata Siudek}
\affiliation{Institute of Space Sciences, ICE-CSIC, Campus UAB, Carrer de Can Magrans s/n, 08913 Bellaterra, Barcelona, Spain}
\affiliation{Instituto Astrofisica de Canarias, Av. Via Lactea s/n, E38205 La Laguna, Spain
}

\author{J.~Aguilar}
\affiliation{Lawrence Berkeley National Laboratory, 1 Cyclotron Road, Berkeley, CA 94720, USA}

\author[0000-0001-6098-7247]{S.~Ahlen}
\affiliation{Physics Dept., Boston University, 590 Commonwealth Avenue, Boston, MA 02215, USA}

\author[0000-0001-9712-0006]{D.~Bianchi}
\affiliation{Dipartimento di Fisica ``Aldo Pontremoli'', Universit\`a degli Studi di Milano, Via Celoria 16, I-20133 Milano, Italy}
\affiliation{INAF-Osservatorio Astronomico di Brera, Via Brera 28, 20122 Milano, Italy}

\author{D.~Brooks}
\affiliation{Department of Physics \& Astronomy, University College London, Gower Street, London, WC1E 6BT, UK}

\author{T.~Claybaugh}
\affiliation{Lawrence Berkeley National Laboratory, 1 Cyclotron Road, Berkeley, CA 94720, USA}

\author[0000-0002-1769-1640]{A.~de la Macorra}
\affiliation{Instituto de F\'{\i}sica, Universidad Nacional Aut\'{o}noma de M\'{e}xico,  Circuito de la Investigaci\'{o}n Cient\'{\i}fica, Ciudad Universitaria, Cd. de M\'{e}xico  C.~P.~04510,  M\'{e}xico}

\author[0000-0002-4928-4003]{Arjun~Dey}
\affiliation{NSF NOIRLab, 950 N. Cherry Ave., Tucson, AZ 85719, USA}

\author{P.~Doel}
\affiliation{Department of Physics \& Astronomy, University College London, Gower Street, London, WC1E 6BT, UK}

\author[0000-0003-4992-7854]{S.~Ferraro}
\affiliation{Lawrence Berkeley National Laboratory, 1 Cyclotron Road, Berkeley, CA 94720, USA}
\affiliation{University of California, Berkeley, 110 Sproul Hall \#5800 Berkeley, CA 94720, USA}

\author[0000-0002-2890-3725]{J.~E.~Forero-Romero}
\affiliation{Departamento de F\'isica, Universidad de los Andes, Cra. 1 No. 18A-10, Edificio Ip, CP 111711, Bogot\'a, Colombia}
\affiliation{Observatorio Astron\'omico, Universidad de los Andes, Cra. 1 No. 18A-10, Edificio H, CP 111711 Bogot\'a, Colombia}

\author{E.~Gaztañaga}
\affiliation{Institut d'Estudis Espacials de Catalunya (IEEC), c/ Esteve Terradas 1, Edifici RDIT, Campus PMT-UPC, 08860 Castelldefels, Spain}
\affiliation{Institute of Cosmology and Gravitation, University of Portsmouth, Dennis Sciama Building, Portsmouth, PO1 3FX, UK}
\affiliation{Institute of Space Sciences, ICE-CSIC, Campus UAB, Carrer de Can Magrans s/n, 08913 Bellaterra, Barcelona, Spain}

\author[0000-0003-3142-233X]{S.~Gontcho A Gontcho}
\affiliation{Lawrence Berkeley National Laboratory, 1 Cyclotron Road, Berkeley, CA 94720, USA}

\author{G.~Gutierrez}
\affiliation{Fermi National Accelerator Laboratory, PO Box 500, Batavia, IL 60510, USA}

\author[0000-0002-6550-2023]{K.~Honscheid}
\affiliation{Center for Cosmology and AstroParticle Physics, The Ohio State University, 191 West Woodruff Avenue, Columbus, OH 43210, USA}
\affiliation{Department of Physics, The Ohio State University, 191 West Woodruff Avenue, Columbus, OH 43210, USA}
\affiliation{The Ohio State University, Columbus, 43210 OH, USA}

\author[0000-0002-6024-466X]{M.~Ishak}
\affiliation{Department of Physics, The University of Texas at Dallas, 800 W. Campbell Rd., Richardson, TX 75080, USA}

\author[0000-0002-0000-2394]{S.~Juneau}
\affiliation{NSF NOIRLab, 950 N. Cherry Ave., Tucson, AZ 85719, USA}

\author[0000-0003-3510-7134]{T.~Kisner}
\affiliation{Lawrence Berkeley National Laboratory, 1 Cyclotron Road, Berkeley, CA 94720, USA}

\author[0000-0001-6356-7424]{A.~Kremin}
\affiliation{Lawrence Berkeley National Laboratory, 1 Cyclotron Road, Berkeley, CA 94720, USA}

\author[0000-0003-1838-8528]{M.~Landriau}
\affiliation{Lawrence Berkeley National Laboratory, 1 Cyclotron Road, Berkeley, CA 94720, USA}

\author[0000-0001-7178-8868]{L.~Le~Guillou}
\affiliation{Sorbonne Universit\'{e}, CNRS/IN2P3, Laboratoire de Physique Nucl\'{e}aire et de Hautes Energies (LPNHE), FR-75005 Paris, France}

\author[0000-0003-4962-8934]{M.~Manera}
\affiliation{Departament de F\'{i}sica, Serra H\'{u}nter, Universitat Aut\`{o}noma de Barcelona, 08193 Bellaterra (Barcelona), Spain}
\affiliation{Institut de F\'{i}sica d’Altes Energies (IFAE), The Barcelona Institute of Science and Technology, Edifici Cn, Campus UAB, 08193, Bellaterra (Barcelona), Spain}

\author[0000-0002-1125-7384]{A.~Meisner}
\affiliation{NSF NOIRLab, 950 N. Cherry Ave., Tucson, AZ 85719, USA}

\author{R.~Miquel}
\affiliation{Instituci\'{o} Catalana de Recerca i Estudis Avan\c{c}ats, Passeig de Llu\'{\i}s Companys, 23, 08010 Barcelona, Spain}
\affiliation{Institut de F\'{i}sica d’Altes Energies (IFAE), The Barcelona Institute of Science and Technology, Edifici Cn, Campus UAB, 08193, Bellaterra (Barcelona), Spain}

\author[0000-0002-2733-4559]{J.~Moustakas}
\affiliation{Department of Physics and Astronomy, Siena College, 515 Loudon Road, Loudonville, NY 12211, USA}

\author{A.~D.~Myers}
\affiliation{Department of Physics \& Astronomy, University  of Wyoming, 1000 E. University, Dept.~3905, Laramie, WY 82071, USA}

\author[0000-0001-9070-3102]{S.~Nadathur}
\affiliation{Institute of Cosmology and Gravitation, University of Portsmouth, Dennis Sciama Building, Portsmouth, PO1 3FX, UK}

\author{C.~Poppett}
\affiliation{Lawrence Berkeley National Laboratory, 1 Cyclotron Road, Berkeley, CA 94720, USA}
\affiliation{Space Sciences Laboratory, University of California, Berkeley, 7 Gauss Way, Berkeley, CA  94720, USA}
\affiliation{University of California, Berkeley, 110 Sproul Hall \#5800 Berkeley, CA 94720, USA}

\author[0000-0001-6979-0125]{I.~P\'erez-R\`afols}
\affiliation{Departament de F\'isica, EEBE, Universitat Polit\`ecnica de Catalunya, c/Eduard Maristany 10, 08930 Barcelona, Spain}

\author{G.~Rossi}
\affiliation{Department of Physics and Astronomy, Sejong University, 209 Neungdong-ro, Gwangjin-gu, Seoul 05006, Republic of Korea}

\author[0000-0002-9646-8198]{E.~Sanchez}
\affiliation{CIEMAT, Avenida Complutense 40, E-28040 Madrid, Spain}

\author{D.~Schlegel}
\affiliation{Lawrence Berkeley National Laboratory, 1 Cyclotron Road, Berkeley, CA 94720, USA}

\author{M.~Schubnell}
\affiliation{Department of Physics, University of Michigan, 450 Church Street, Ann Arbor, MI 48109, USA}
\affiliation{University of Michigan, 500 S. State Street, Ann Arbor, MI 48109, USA}

\author[0000-0002-6588-3508]{H.~Seo}
\affiliation{Department of Physics \& Astronomy, Ohio University, 139 University Terrace, Athens, OH 45701, USA}

\author[0000-0002-3461-0320]{J.~Silber}
\affiliation{Lawrence Berkeley National Laboratory, 1 Cyclotron Road, Berkeley, CA 94720, USA}

\author[0000-0003-1704-0781]{G.~Tarl\'{e}}
\affiliation{University of Michigan, 500 S. State Street, Ann Arbor, MI 48109, USA}

\author{B.~A.~Weaver}
\affiliation{NSF NOIRLab, 950 N. Cherry Ave., Tucson, AZ 85719, USA}

\author[0000-0001-5381-4372]{R.~Zhou}
\affiliation{Lawrence Berkeley National Laboratory, 1 Cyclotron Road, Berkeley, CA 94720, USA}

\author[0000-0002-6684-3997]{H.~Zou}
\affiliation{National Astronomical Observatories, Chinese Academy of Sciences, A20 Datun Rd., Chaoyang District, Beijing, 100012, P.R. China}

\begin{abstract}
We investigate the relationships between the cool circumgalactic medium (CGM), traced by \caii absorption lines, and galaxy properties at $z<0.4$ using $\sim900{,}000$ galaxy--quasar pairs within $200\,\rm kpc$ from the Year 1 data of the Dark Energy Spectroscopic Instrument (DESI). This large data set enables us to obtain composite spectra with sensitivity reaching to the $\rm m\AA$ level and to explore the \caii absorption as a function of stellar mass, star formation rate (SFR), redshift, and galaxy types, including active galactic nuclei (AGNs). Our results show a positive correlation between the absorption strength and stellar mass of star-forming galaxies with $\langle W_{0}^{\rm Ca\ II}\rangle \propto M_{*}^{0.5}$ over 3 orders of magnitude in stellar mass from $\sim 10^{8}$ to $10^{11} \, M_{\odot}$, while such a mass dependence is weaker for quiescent galaxies. At a fixed mass, \caii absorption is stronger around star-forming galaxies than quiescent ones especially within impact parameters $<30\,\rm kpc$. Among star-forming galaxies, the \caii absorption further correlates with SFR, following $\propto \mathrm{SFR^{0.3}}$. However, in contrast to the results at higher redshifts, stronger absorption is not preferentially observed along the minor axis of star-forming galaxies, indicating a possible redshift evolution of CGM dynamics resulting from galactic feedback. Moreover, no significant difference between the properties of the cool gas around AGNs and galaxies is detected. Finally, we measure the absorption profiles with respect to the virial radius of dark matter halos and show that the total \caii mass in the CGM is comparable to the Ca mass in the ISM of galaxies.

\end{abstract}


\section{Introduction} \label{sec:intro}
The circumgalactic medium (CGM) contains signatures of complex physical processes that regulate gas flowing in and out of galaxies and drive galaxy evolution \citep[see][for reviews on the topic of the CGM]{cgm, araa20, cgmsim}. Therefore, observed properties of the CGM and its connection to the properties of galaxies have been considered as crucial constraints on the models of galaxy evolution, especially the subgrid models implemented for the feedback processes driven by the explosions of massive stars and the activities of supermassive black holes \citep[e.g.,][]{ford16, liang16, angles17, oppenheimer18}. To probe the properties of the CGM,  absorption line spectroscopy has been widely used \citep[e.g.,][]{bahcall69, Bergeron86, chen10, nielsen13, tumlinson13, lan14, zhu14, anand21, bouche25}. By observing the spectra of background sources intercepting the CGM of galaxies, one can detect absorption line features induced by various transitions of elements, and investigate the properties of the circumgalactic gas, such as the radial distribution, kinematics, abundance, metallicity, and gas density, and explore how they correlate with the properties of the galaxies \citep[e.g.,][]{zych09,zhu14,lan14,nielsen17, pointon19,lan20, anand21, lehner24, nateghi24,cherrey25}. 

Multiwavelength observations are required to investigate the galaxy--CGM connections at different redshifts owing to the accessibility of 
strong absorption-line transitions. At $z<0.3$, UV spectrographs, COS and STIS \citep{cos, stis}, on the Hubble Space Telescope have been used to detect absorption species, such as \ion{H}{1}, \ion{O}{1}, \ion{O}{6}, \ion{C}{2}, \ion{C}{4}, \ion{Si}{2}, and \ion{Si}{3}, tracing gas in various physical conditions \citep[e.g.,][]{tumlinson13, werk13, werk14, bordoloi14, liang14, borthakur15, borthakur16, werk16, prochaska17, chen20, tchernyshyov23, zheng24}. These measurements have led to new explorations of the nature and origins of the multiphase CGM. However, given the sensitivity of the current instruments, only bright quasars can yield UV spectra with sufficient sensitivity for absorption-line studies \citep[e.g.,][]{tumlinson13}. This in turn limits the number of galaxy--quasar pairs and the parameter space of galaxy properties available for galaxy--CGM correlation analysis. On the other hand, at $z\gtrsim 0.4$, optical spectra can be used to investigate the connection between galaxies and the cool CGM traced by \mgii absorption lines, one of the key accessible absorption species at optical wavelengths \citep[e.g.,][]{nestor05, chen10, nielsen13, zhu13mg, anand21, huang21, chen23, guha24, zou24, bouche25}. With large optical spectroscopic data sets provided by sky surveys, such as the Sloan Digital Sky Survey (SDSS; \citealt{york00}) and the Dark Energy Spectroscopic Instrument (DESI; \citealt{desi13}), statistical samples, consisting of thousands of galaxy--quasar pairs, have been built and used to measure the properties of the cool CGM as a function of galaxy properties, including stellar mass, star formation rate (SFR) and active galactic nucleus (AGN) activity, across redshifts \citep[e.g.,][]{zhu13mg, lan14, lan20, anand21, chang24, wu24}, revealing correlations between the cool CGM and galaxy properties. However, observational measurements of these correlations at low redshifts are still limited, and therefore whether the relationships between galaxies and the cool CGM detected at $z\gtrsim0.4$ exist at lower redshifts remains to be addressed. 

At $z<0.4$, while the majority of absorption transitions are in the UV wavelength regions, the \caii doublet H ($3969.59 \textup{~\AA}$) and K ($3934.78 \textup{~\AA}$) lines, which trace cool gas at $T\sim 10^4 -10^{4.5}\,\mathrm{K}$---similar to \mgii and Ly$\alpha$ absorbers---are accessible in optical wavelengths. Despite the fact that \caii lines are weak---
which restricts individual detections to high column density systems \citep[e.g.,][]{wild06, hewett07, nestor08, zych09, richter11, sardane14, sardane15, zou18}---one can detect and measure their absorption strengths around galaxies via high S/N composite spectra obtained by statistically combining many individual spectra \citep[e.g.,][]{steidel10, bordoloi11, zhu14, pieri14, lan18, wu24, changlan25, chen25}. With this approach, \citet{zhu13} demonstrated that \caii absorption features can be detected and used to explore the cool CGM properties around galaxies with $\sim 10^{10} \, M_{\odot}$ at $z\sim0.1$ using galaxy and quasar spectroscopic data sets from the SDSS. 
Combining this method with spectroscopic data sets covering a wider range of galaxy properties, one can further obtain a more comprehensive characterization of the galaxy--cool CGM connections at low redshifts and compare with such connections observed at higher redshifts.  

In this work, we utilize the large spectroscopic data provided by the DESI survey \citep{desi1, desi2} to measure the properties of the cool CGM traced by \caii as a function of galaxy properties, including SFR, stellar mass, redshift, and azimuthal angle, covering a parameter space that has rarely been probed previously. We describe our sample and the analysis procedure to obtain composite spectra in Section \ref{sec:data}. In Section \ref{sec:results}, we present our results. In Section \ref{sec:discussion}, we compare our results with previous studies and discuss the implications of our findings. Finally, we conclude our results in Section \ref{sec:conclusion}. Throughout this paper, we adopt a flat $\Lambda$CDM model with $h = 0.7$ and $\Omega_m=0.3$.

\section{Data analysis} \label{sec:data}

\subsection{DESI galaxy and quasar catalogs}
The DESI project is designed to measure the expansion rate of the Universe across cosmic time by using galaxies, quasars, and gas as tracers of large-scale structure at different redshifts \citep{desi13, desioverview}. To this end, a dedicated spectroscopic observational program has been developed utilizing an instrument consisting of 10 spectrographs and 5000 fibers \citep{focalplane, corrector}. DESI primarily obtains spectra of four types of extragalactic sources at different redshift ranges, including galaxies that can be observed during the bright time targeted by the Bright Galaxy Survey ($z<0.6$; \citealt{bgs}), luminous red galaxies ($0.4<z<1.0$; \citealt{lrg}), emission-line galaxies ($0.6<z<1.6$; \citealt{elg}) and quasars ($z<3$; \citealt{qso}). These sources are selected based on the images of the DESI Legacy Imaging Surveys \citep{dey19}. The wavelength coverage of DESI spectra is from 3600 to 9800\textup{~\AA} with the spectral resolution varying from $R \sim 2000$ to 5000 for the blue camera to the NIR camera. The raw spectroscopic data are processed with a pipeline developed by \cite{guy23}. In addition, an automatic pipeline, called \textsc{redrock}\footnote{\url{https://github.com/desihub/redrock}}, is developed and used for determining the redshifts of the sources from the spectra (\citealt{Redrock.Bailey.2024}; see also \citealt{anand24}). In order to maximize the efficiency of the survey, dedicated algorithms and codes have also been developed for target selection \citep{myers23}, fiber assignments \citep{FBA.Raichoor.2024} and survey operations \citep{schlafly23}. Prior to the main spectroscopy survey, DESI has undergone the ``Survey Validation (SV)" \citep{sv} to validate the performance of the target selections. The five-month spectroscopic data from Survey Validation was released in 2023 June as part of Early Data Release \citep{edr}.

\begin{figure}
\centering
\includegraphics[width=\columnwidth]{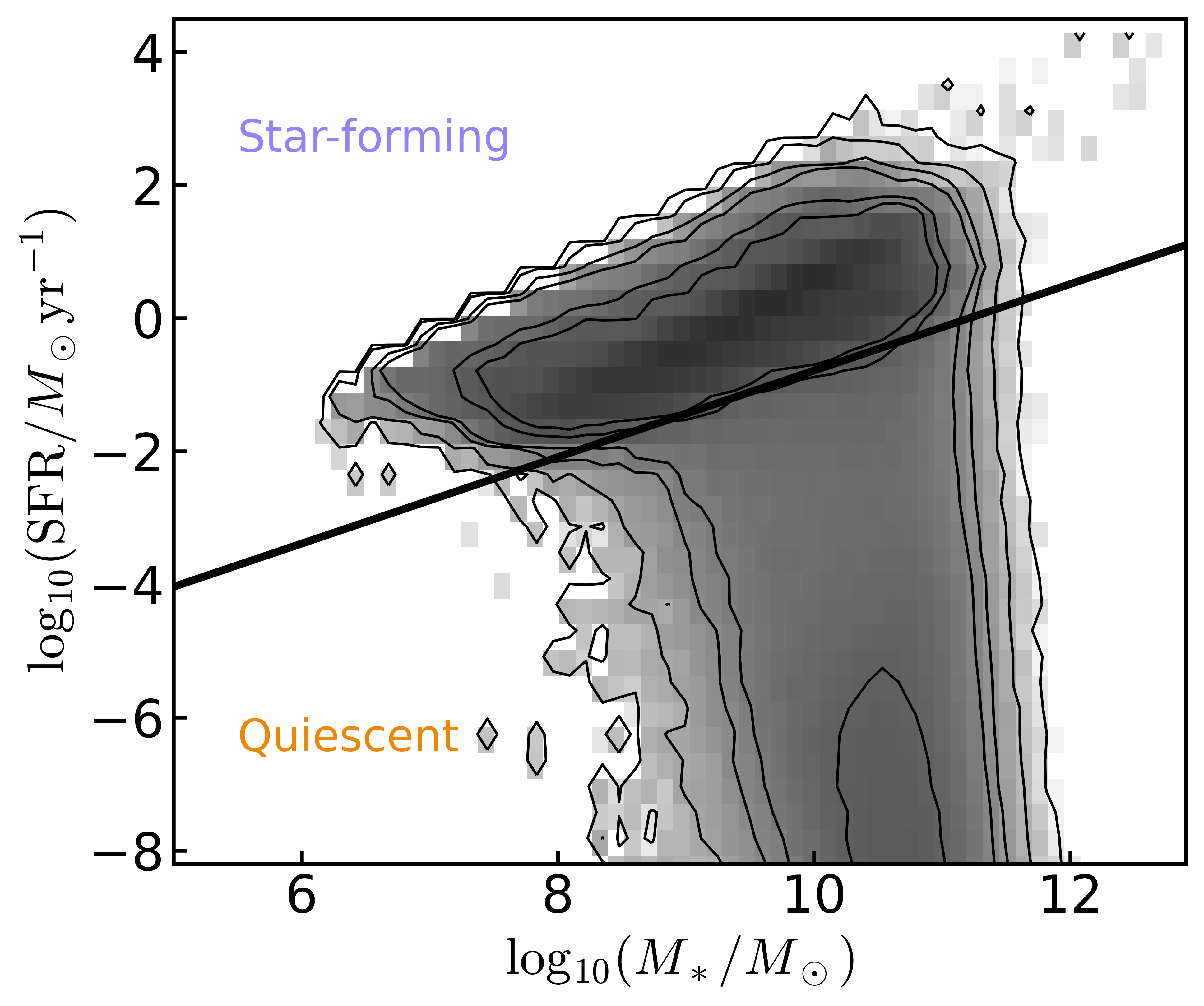}
\caption{SFR--mass distribution of the foreground galaxies. The solid line divides the samples into star-forming and quiescent galaxies. The contours mark the levels of numbers at 10, 100, 500, 1000, 5000, and 8000.}
\label{fig:sfr_div}
\end{figure}

In this work, we use the Year 1 data set of the DESI survey \citep{desi_i}, which has yielded new measurements for constraining the expansion history of the Universe from $z = 0$ to 2 \citep{desi_ii, desi_iii, desi_iv, desi_v, desi_vi, desi_vii, desi_dr2_i, desi_dr2_ii}. More specifically, we construct our foreground galaxy and background source samples with the following two components:
\begin{itemize}
    \item \textbf{Bright Galaxy Survey (BGS)}: We use galaxies with $0.02<z_\mathrm{BGS}<0.4$ observed in the bright galaxy survey as our foreground galaxies. The BGS sample consists of two selections, BGS bright and BGS faint samples. The BGS bright sample is selected with $r<19.5$ mag, and the BGS faint sample is selected with magnitude and color cuts \citep{bgs}. Moreover, following the criteria from \cite{bgs} and \cite{vi}, we select galaxies with $\Delta \chi^2 \ge 40$ (the difference of $\chi^2$ between the second and the first best-fit spectral models from \textsc{redrock}). 

    Besides the redshift information, we use the physical properties of the BGS galaxies, including the stellar mass ($M_{*}$) and SFR, provided by \citet{gosia24}. The authors use the CIGALE package \citep{cigale} for spectral energy distribution fits to
    observed magnitudes in the $g$, $r$, $z$, WISE 1 and WISE 2 bands extracted by the \textsc{Tractor} algorithm \citep{lang16}. The final fitting results exclude sources with low signal-to-noise ratio in photometry and poor fitting quality. For details, we refer the readers to \citet{gosia24}. 

    Figure~\ref{fig:sfr_div} shows the stellar mass and SFR distribution of the BGS galaxy sample used in this analysis. To separate star-forming galaxies and quiescent galaxies, we adopt the following relation (black solid line) from \cite{zhu13}, which was derived from a sample of galaxies with a mean redshift $z \sim 0.1$,
    \begin{equation}
        \log_{10}{\frac{\mathrm{SFR}}{M_\odot\, \mathrm{yr^{-1}}}} = -0.79 + 0.65\, \left(\log_{10}{\frac{M_*}{M_\odot}}-10.0 \right). \label{eqn: masssfr}
    \end{equation}
    As shown in the figure, the BGS sample includes star-forming galaxies with $M_{*}$ down to $10^{7}-10^{8}\,M_\odot$ allowing us to probe the CGM of galaxies with mass across 3--4 orders of magnitude.
    The median mass uncertainty of the sample is 0.12 dex. For SFR, the typical uncertainties are 0.3 dex for star-forming galaxies and 8.1 dex for quiescent galaxies. We note that the reported uncertainty is in logarithmic scale. For quiescent galaxies with low estimated SFR ($\log_{10} \mathrm{SFR}/M_{\odot} \rm yr^{-1}\sim-7$), the corresponding uncertainty in linear scale is $\sim10^{-6} M_{\odot}\rm yr^{-1}$. This precision is sufficient for classifying galaxies into star-forming and quiescent ones.
    
    In addition, we construct an AGN sample by selecting sources located in the AGN and composite regions \citep{kewley06} of the BPT diagram \citep{bpt}. For this selection, we require a signal-to-noise ratio greater than 2 for the $\rm H\alpha$, $\rm H\beta$, [\ion{O}{3}] 5007, and [\ion{N}{2}] 6584 emission lines. The flux measurements for these lines are provided by the FastSpecFit\footnote{\url{https://fastspecfit.readthedocs.io/en/latest/}} Spectral Synthesis and Emission-Line Catalog (FastSpecFit 2.1; \citealt{fastspecfit}).

    \item \textbf{Quasars}: For the background sources, we use the quasar sample that is selected with a random forest algorithm combining magnitude and color information \citep{qso}. Similar to the BGS sample selection, we apply $\Delta \chi^2 \ge 20$ to obtain sources with robust \textsc{redrock} redshift measurements \citep{qso_chi2}. 
\end{itemize}

With the galaxy and quasar samples, we construct the galaxy--quasar pair sample by requiring (1) $z_\mathrm{QSO} - z_\mathrm{BGS} > 0.1$ and (2) $\left(z_\mathrm{BGS}+1\right)\times3800\,\mathrm{\AA} > \left(z_\mathrm{QSO}+1\right)\times 1216\,\mathrm{\AA}$, ensuring that the \caii lines fall outside the Ly$\alpha$ forest of the background quasar to avoid contamination.
This yields a sample with impact parameter $r_p \le 200\,\mathrm{kpc}$ having 855,703 galaxy--quasar pairs with 623,053 unique BGS galaxies and 496,756 unique QSOs. The median redshift of the unique BGS galaxies is $\langle z \rangle = 0.1712$. 

\begin{figure}\centering
\includegraphics[width=\columnwidth]{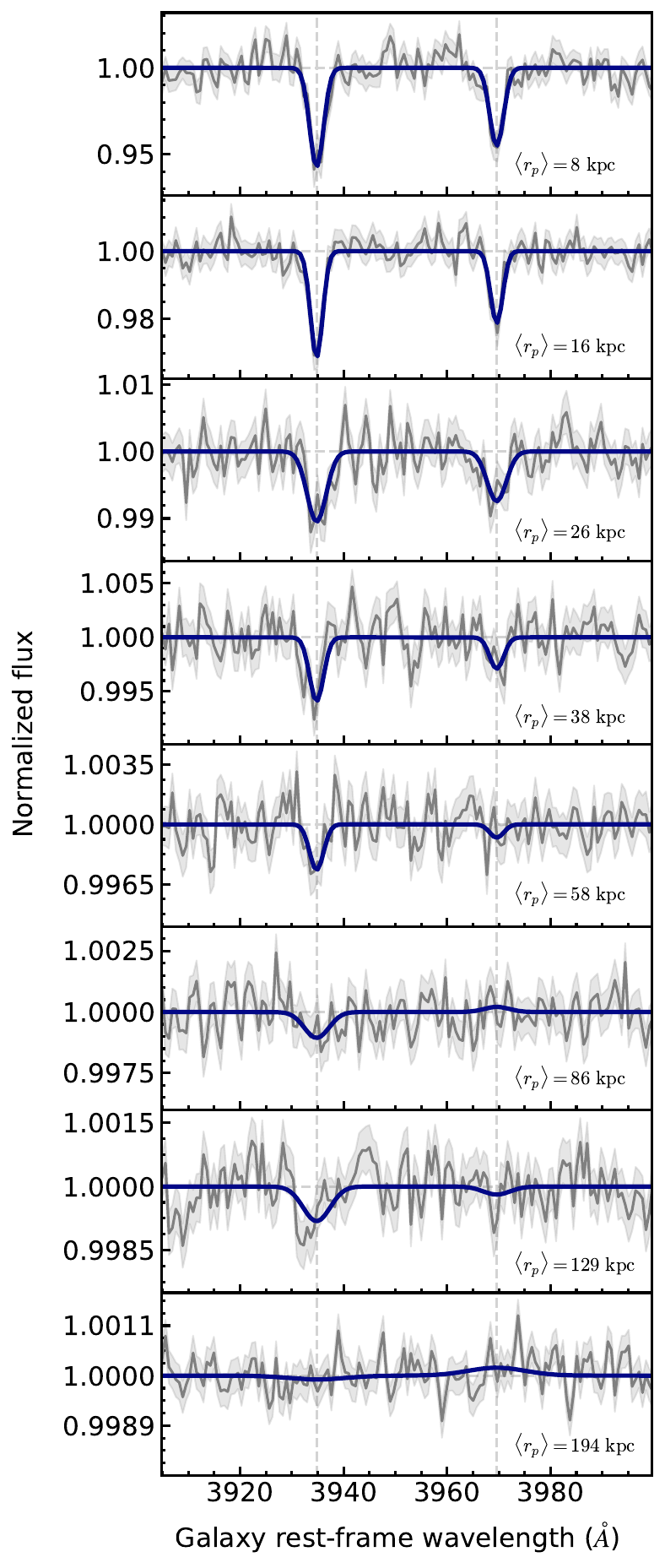}
\caption{Examples of composite spectra as a function of impact parameters. The shades underneath are the associated uncertainties via bootstrapping. The solid blue lines are the best-fit absorption profiles of \caii K and H. The vertical dashed lines are the central wavelengths of K and H lines. The median impact parameter is displayed in the lower right corner. Note that the scale of the y-axis changes as a function of impact parameters.}
\label{fig:stacked_spec}
\end{figure}

\subsection{Constructing composite spectra}
To extract the absorption lines from the quasar spectra, we first employ a dimensional reduction technique, called nonnegative matrix factorization (NMF; \citealt{lee1999learning}), with code\footnote{\url{https://github.com/guangtunbenzhu/NonnegMFPy}} developed by \cite{zhu13mg} and \cite{zhu16}. With NMF, we normalize the observed quasar spectrum by the reconstructed continuum, derived from a basis set of NMF eigenspectra built from SDSS DR7 quasars \citep{Schneider10, zhu13mg}, to remove the spectral features intrinsic to the quasar. To further remove fluctuations not captured by the NMF method, we obtain a smooth continuum by applying a median filter with a window size of 71 pixels ($\sim 56\,\rm\AA$) to the NMF normalized spectrum and divide the NMF normalized spectrum by the smooth continuum. The window size is empirically chosen to be large enough to ensure sensitivity to the continuum rather than the absorption itself, which spans about 5 pixels (a full width at half maximum of $\sim \mathrm{3.5\,\AA}$). We then shift the normalized quasar spectra to the rest frame of the foreground galaxies, and resample them onto a common wavelength grid with a resolution matching that of the DESI spectra when the doublet line is redshifted to $z=0.2$ ($\sim \mathrm{0.67\,\AA}$). We obtain composite spectra with a median estimator to mitigate the contributions from spectral outliers that occasionally occur. Finally, to account for a subpercent of zero-point offset, the final composite spectrum is obtained by normalizing the composite spectrum by the median value of the pixels within 3900 and 4100 $\rm \AA$ with the pixels within 5 $\rm \AA$ of \caii K and H lines being masked. The uncertainties of the median composite spectra are estimated by bootstrapping the samples 500 times. 

To measure the \caii rest equivalent width, we fit the composite spectra with two Gaussian profiles, fixing the central wavelengths at 3934.78\textup{~\AA} (\caii K) and 3969.59\textup{~\AA} (\caii H). The amplitudes and line widths are free parameters, but the widths of both lines are enforced to be the same. Throughout this work, we report the sum of the rest equivalent widths of the \caii K and H lines, denoted as $W_{0}^{\rm K+H} (\mathrm{Ca\ II})$. 
The uncertainties of the rest equivalent width are based on fitting the 500 bootstrapped samples and taking the standard deviation. Figure~\ref{fig:stacked_spec} shows examples of composite spectra as a function of impact parameter; the best-fit profiles are shown by the blue lines.

\begin{figure*}[ht!]
\centering
\includegraphics[width=\textwidth]{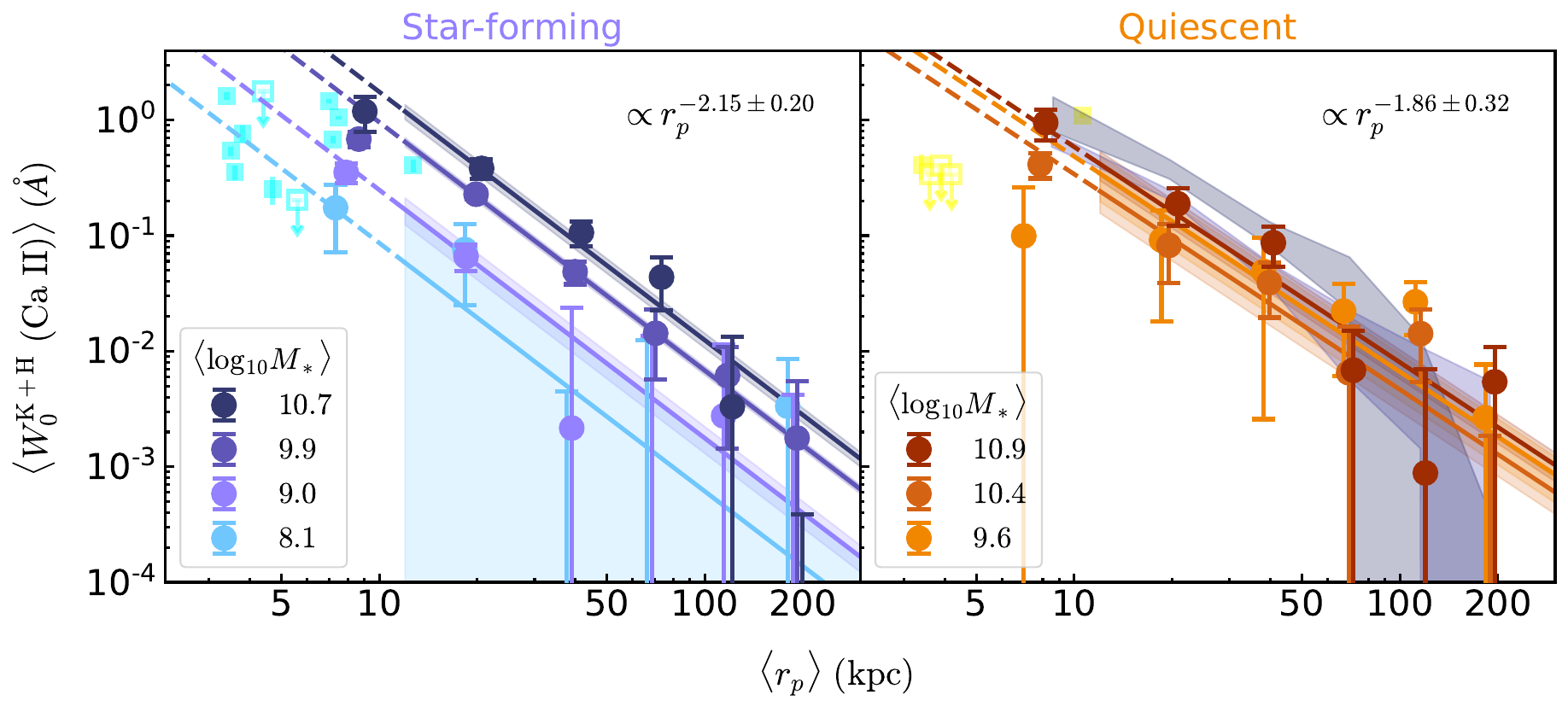}
\caption{Dependence of the \caii K and H equivalent width on stellar masses as a function of impact parameters. Left: star-forming galaxies. Right: quiescent galaxies. The median stellar masses are listed in the lower left. The blue shaded bands in the right panel correspond to the two highest mass bins for star-forming galaxies in the left panel. The cyan and yellow square data points are the K line measurements from \cite{rubin22} multiplied by a factor of 1.69 to compare with our doublet measurements.}
\label{fig:4mass}
\end{figure*}

\begin{figure}[ht!]
\centering
\includegraphics[width=\columnwidth]{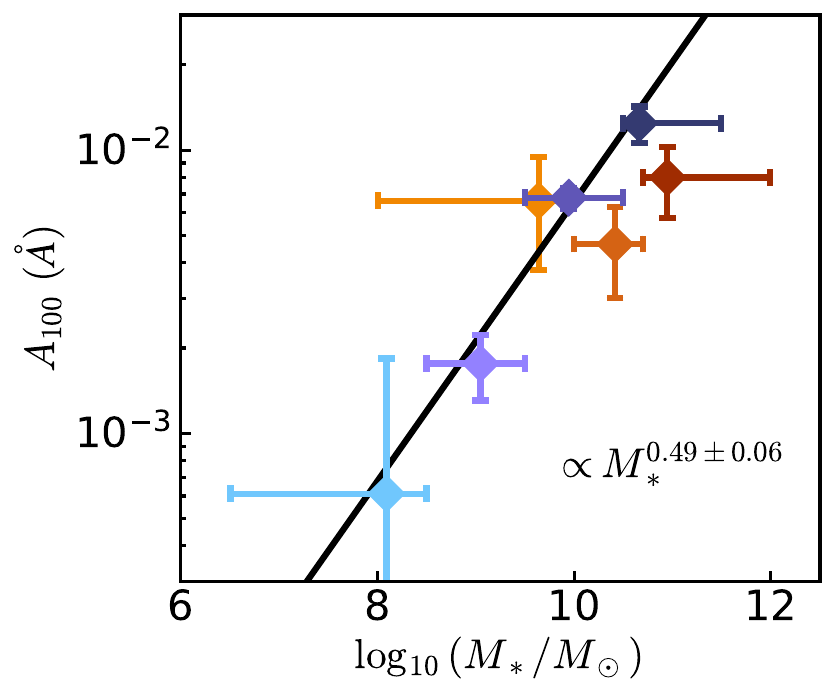}
\caption{Best-fit amplitudes $A_{100}$ as a function of stellar masses. The data points are plotted with their median stellar masses and respective ranges. The black line is the best-fit power law for the star-forming galaxies.}
\label{fig:4mass_power}
\end{figure}

\section{Results} \label{sec:results}

\subsection{Dependencies on Galaxy Properties} \label{sec:massfr}
With the large number of galaxy--quasar pairs across a wide range of galaxy properties, we first investigate the \caii rest equivalent width as a function of impact parameter and stellar mass, comparing star-forming and quiescent galaxies. The results are shown in Figure~\ref{fig:4mass} with the left and right panels indicating the \caii rest equivalent width as a function of impact parameter around star-forming and quiescent galaxies, respectively, with different stellar masses indicated by the colors. 
The number of pairs is summarized in Table~\ref{tab:mass}. We find several trends, which are described in the following:

\begin{deluxetable*}{ccccccc}
\tablecaption{Mass Bins and Best-fit Values for Star-forming and Quiescent Galaxies}
\label{tab:mass}
\renewcommand{\tabcolsep}{4.mm}
\tablehead{{$x$-variable} & {Galaxy Type} & \colhead{$z$\tablenotemark{\scriptsize{a}}} & \colhead{$\log_{10}{(M_*/M_\odot)}$\tablenotemark{\scriptsize{a}}}  & \colhead{$A_{100}\ |\ A_{0.5\rm vir}\ \ (\rm m\AA)$} & \colhead{$\alpha\ |\ \eta$} & \colhead{Number of Pairs\tablenotemark{\scriptsize{b}}}} \startdata
\hline
\multirow{22}{*}{$r_p$} & \multirow{11}{*}{Star-forming} & \multirow{4}{*}{$0.08_{-0.06}^{+0.32}$} & $\phantom{0}8.1_{-1.6}^{+0.4}$  & $\phantom{0}0.61\pm1.23$ & \multirow{4}{*}{$-2.15\pm0.20$} & 139,946 \\
& & & $\phantom{0}9.0_{-0.5}^{+0.5}$ & $\phantom{0}1.76\pm0.46$ & & 263,731 \\
& & & $\phantom{0}9.9_{-0.4}^{+0.6}$ & $\phantom{0}6.79\pm0.57$ & & 285,305 \\
& & & $10.7_{-0.2}^{+0.8}$ & $12.43\pm1.81$ & & \phantom{0}47,469 \\
\cline{3-7}
& & \multirow{3}{*}{$0.07_{-0.05}^{+0.08}$
} & $\phantom{0}8.9_{-0.6}^{+0.5}$ & $\phantom{0}1.50\pm0.47$ & & 261,176 \\
& & & $\phantom{0}9.8_{-0.4}^{+0.6}$ & $\phantom{0}4.88\pm0.78$ & & 166,954 \\
& & & $10.6_{-0.2}^{+0.6}$ & $13.85\pm3.74$ & & \phantom{0}18,175 \\
\cline{3-7}
& & \multirow{2}{*}{$0.22_{-0.07}^{+0.18}$ } & $10.0_{-0.6}^{+0.4}$  & $\phantom{0}6.33\pm2.60$ & & 130,437 \\
& & & $10.6_{-0.2}^{+0.6}$ & $\phantom{0}9.43\pm6.73$ & & \phantom{0}48,104 \\
\cline{2-7}
& \multirow{7}{*}{Quiescent} & \multirow{3}{*}{$0.15_{-0.13}^{+0.25}$} & $\phantom{0}9.6_{-1.6}^{+0.4}$ & $\phantom{0}6.62\pm2.85$ & \multirow{3}{*}{$-1.86\pm0.32$} & \phantom{0}61,976 \\
& & & $10.4_{-0.4}^{+0.3}$ & $\phantom{0}4.66\pm1.65$ & & 134,889 \\
& & & $10.9_{-0.2}^{+1.0}$ & $\phantom{0}8.01\pm2.25$ & & 116,110 \\
\cline{3-7}
& & \multirow{2}{*}{$0.10_{-0.08}^{+0.07}$ } & $10.3_{-0.8}^{+0.7}$ & $\phantom{0}7.94\pm0.82$ & & 142,711 \\
& & & $11.1_{-0.1}^{+0.6}$ & $11.68\pm1.60$ & & \phantom{0}13,032 \\
\cline{3-7}
& & \multirow{2}{*}{$0.25_{-0.08}^{+0.15}$ } & $10.7_{-0.5}^{+0.3}$ & $\phantom{0}4.63\pm1.83$ & & \phantom{0}91,412 \\
& & & $11.2_{-0.2}^{+0.5}$ & $14.88\pm3.64$ & & \phantom{0}34,505 \\
\cline{2-7}
& AGN & $0.07_{-0.05}^{+0.33}$ & $10.2_{-3.2}^{+1.3}$ & $\phantom{0}4.80\pm1.27$ & $-2.49\pm1.29$ & \phantom{0}39,559 \\
\noalign{\hrule height.8pt}
\multirow{6}{*}{$r_p/r_\mathrm{vir}$} & \multirow{3}{*}{Star-forming} & \multirow{3}{*}{$0.20_{-0.18}^{+0.20}$} & $\phantom{0}8.8_{-2.3}^{+0.4}$ & $\phantom{0}4.05\pm0.93$ & \multirow{3}{*}{$-2.29\pm0.24$} & \phantom{0}69,724 \\
& & & $\phantom{0}9.8_{-0.6}^{+0.4}$ & $\phantom{0}6.53\pm0.61$ & & 154246 \\
& & & $10.6_{-0.4}^{+0.9}$ & $\phantom{0}6.21\pm0.74$ & & 143,798 \\
\cline{2-7}
& \multirow{3}{*}{Quiescent} & \multirow{3}{*}{$0.22_{-0.20}^{+0.18}$} & $\phantom{0}9.7_{-1.7}^{+0.3}$ & $\phantom{0}7.25\pm4.48$ & \multirow{3}{*}{$-1.90\pm0.47$} & \phantom{0}28,399 \\
& & & $10.5_{-0.5}^{+0.2}$ & $\phantom{0}5.13\pm1.29$ & & 143,912 \\
& & & $11.0_{-0.3}^{+1.0}$ & $\phantom{0}1.08\pm0.62$ & & 364,133
\enddata
\tablenotetext{a}{Median quantities of pairs with the ranges of lower and upper limits.}
\tablenotetext{b}{$r_p<228\rm \, kpc$ for the measurements as a function of $r_p$, and $r_p<500\rm \, kpc$ for the measurements as a function of $r_p/r_\mathrm{vir}$.}
\end{deluxetable*} 
\textbf{Mass dependence:} The absorption strengths of \caii correlate with the masses of star-forming galaxies and quiescent galaxies differently. The $W_{0}^{\rm K+H}$ exhibits an increasing trend with the stellar mass of star-forming galaxies, whereas this trend is much weaker for quiescent galaxies, albeit with large uncertainties. To further quantify these correlations, we fit the $W_{0}^{\rm K+H}$ with a simple power-law relation,
\begin{equation}
        \langle W_0 \rangle(r_p) = A_{100} \left(\frac{r_p}{100\ \mathrm{kpc}}\right)^\alpha, \label{eqn:powerlaw}
\end{equation}
where $A_{100}$ is the equivalent width around galaxies at $r_p=100\,\mathrm{kpc}$. We first perform a global fitting to obtain the best-fit $\alpha$ parameter for each type of galaxy, assuming that the radial variations for the same type of galaxies are the same. We then fix the $\alpha$ parameters and obtain the best-fit amplitude for each stellar mass bin. Note that we do not include measurements within 10 kpc as the sightlines are expected to also intersect the ISM of foreground galaxies, affecting our interpretations for the CGM. The best-fit profiles are shown by the color solid lines in Figure~\ref{fig:4mass} with the best-fit parameter values listed in Table~\ref{tab:mass}. The best-fit amplitudes, $A_{100}$, as a function of stellar mass are shown in Figure~\ref{fig:4mass_power} with the colors representing the corresponding subsamples as in Figure~\ref{fig:4mass}. The results demonstrate that \caii absorption strengths around star-forming galaxies at $\sim 100\,\mathrm{kpc}$ exhibit a mass dependence that can be described by 
\begin{equation}
  A_{100} = C\left(\frac{M_{*}}{10^{10} \, M_{\odot}}\right)^{\beta},
  \label{eq:mass_depedence}
\end{equation}
with best-fit $\beta$ parameter values being $0.49 \pm 0.06$, and $C=6.57\pm0.47\ \rm m\AA$.
On the other hand, the absorption strengths around quiescent galaxies do not depend significantly on stellar mass. 

\textbf{Galaxy type dependence:} 
In addition to the difference of mass dependence, for massive galaxies ($>10^{10} M_{\odot}$) with similar mass, at $r_{p}<30 \rm \, kpc$, the absorption strengths around star-forming galaxies are higher than those around quiescent galaxies. By way of illustration, in the right panel of Figure~\ref{fig:4mass}, we show the \caii measurement star-forming galaxies with $10^{10.7} M_{\odot}$ and $10^{9.9} M_{\odot}$. Comparing with the two highest stellar mass bins of quiescent galaxies, the \caii absorption within 50 kpc around $10^{10.7} M_{\odot}$ and $10^{9.9} M_{\odot}$ star-forming galaxies is twice as strong as the absorption around $10^{10.9} M_{\odot}$ and $10^{10.4} M_{\odot}$ quiescent galaxies, despite the star-forming galaxies having lower masses. This indicates that the cool gas distributes around star-forming and quiescent galaxies differently
when one controls for stellar mass.

\begin{figure}[ht!]
\centering
\includegraphics[width=\columnwidth]{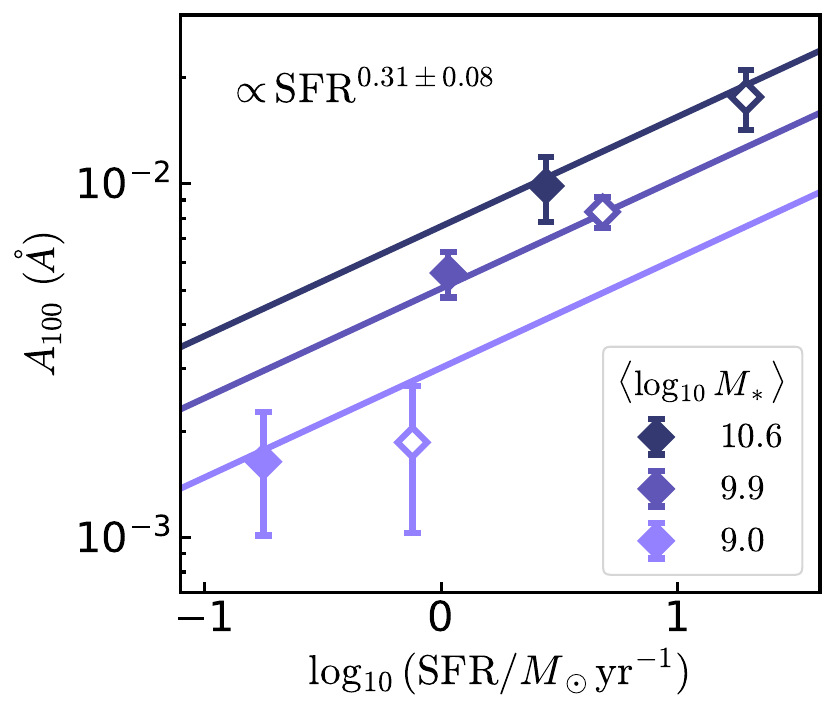}
\caption{Best-fit amplitudes $A_{100}$ as a function of SFR for star-forming galaxies. The solid and open points represent the low- and high-SFR groups, respectively, with matched median stellar masses. The lines correspond to the best-fit power law obtained from simultaneous fitting.}
\label{fig:sfr}
\end{figure}

\textbf{SFR dependence:} We further examine the dependence of \caii absorptions on star formation activity by dividing star-forming galaxies into low- and high- SFR groups using the equation: $
        \log_{10}{(\mathrm{SFR}/M_\odot\, \mathrm{yr^{-1}})} = 0.5 + 0.95\, \left[\log_{10}{(M_*/M_\odot)}-10.0 \right]. $
The samples for both SFR groups are then divided into three mass bins with matched median masses of $10^{9}$, $10^{9.9}$, and $10^{10.6}\,M_\odot$. We follow the previous steps to obtain the \caii absorption strength as a function of impact parameter, fitting the trend globally with $\alpha=-2.15$. The fitted $A_{100}$ values are shown in Figure~\ref{fig:sfr}. By simultaneously fitting $A_{100}$ and SFR across the three mass bins, we find a positive correlation between \caii absorption strength and SFR, following 
\begin{equation}
    A_{100} = C\left(\frac{M_{*}}{10^{10} \, M_{\odot}}\right)^{\beta} \left(\frac{\mathrm{SFR}}{ M_{\odot}\mathrm{yr^{-1}}}\right)^{\gamma},
\end{equation}
where $C=5.22\pm0.61\,\mathrm{m\AA}$, $\beta=0.25\pm0.10$, and $\gamma = 0.31\pm0.08$.
This result further supports the connection between the cool CGM and the star formation activity of host galaxies.

\begin{figure}[htb!]
\centering
\includegraphics[width=\columnwidth]{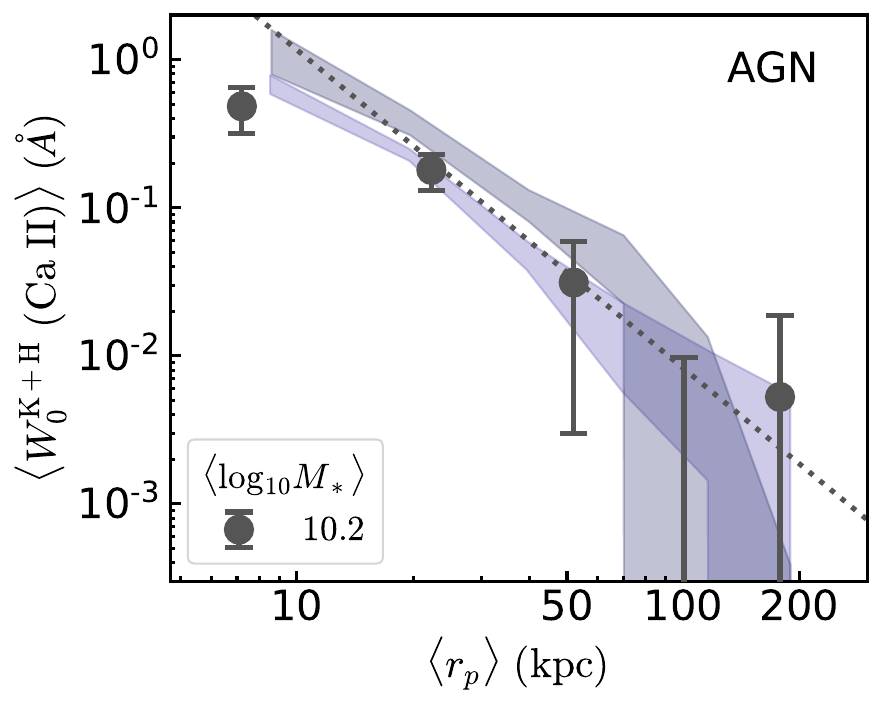}
\caption{Equivalent width of AGN samples as a function of impact parameters. The shaded bands correspond to the two highest mass bins ($10^{9.9}$ and $10^{10.7}\,M_\odot$) from star-forming galaxies in Figure~\ref{fig:4mass}. The dotted line is calculated using Equation~\ref{eq:mass_depedence} with $M_*=10^{10.2}M_\odot$.}
\label{fig:AGN}
\end{figure}

\textbf{AGN dependence:} 
The large DESI data set enables us to compare the \caii absorption signal around  galaxies with and without an AGN, as identified with the BPT diagnosis \citep{kewley06}. The active galaxies sample includes sources in the composite and AGN regions of the BPT diagram, with a median stellar mass of $\sim 10^{10.2} \, M_{\odot}$ and a range from $10^{7}$ to $10^{11.5}\, M_{\odot}$. The results are shown in Figure~\ref{fig:AGN} with the black data points showing the measurements around AGNs and the blue bands showing the measurements around star-forming galaxies with median masses of $10^{9.9}$ and $10^{10.7} \, M_{\odot}$ for comparison. In addition, the dotted line in Figure~\ref{fig:AGN} shows the expected \caii profile for $10^{10.2} \, M_{\odot}$ star-forming galaxies based on the best-fit relation (Equation~\ref{eq:mass_depedence}). 

The results show that the \caii rest equivalent width around AGNs is consistent with that around star-forming galaxies with similar masses, indicating that current AGN activity does not significantly affect the properties of the cool gas traced by \caii around galaxies. This result is consistent with previous studies, which will be discussed in Section~\ref{sec:discussion}.

\begin{figure*}[ht!]
\centering
\includegraphics[width=\textwidth]{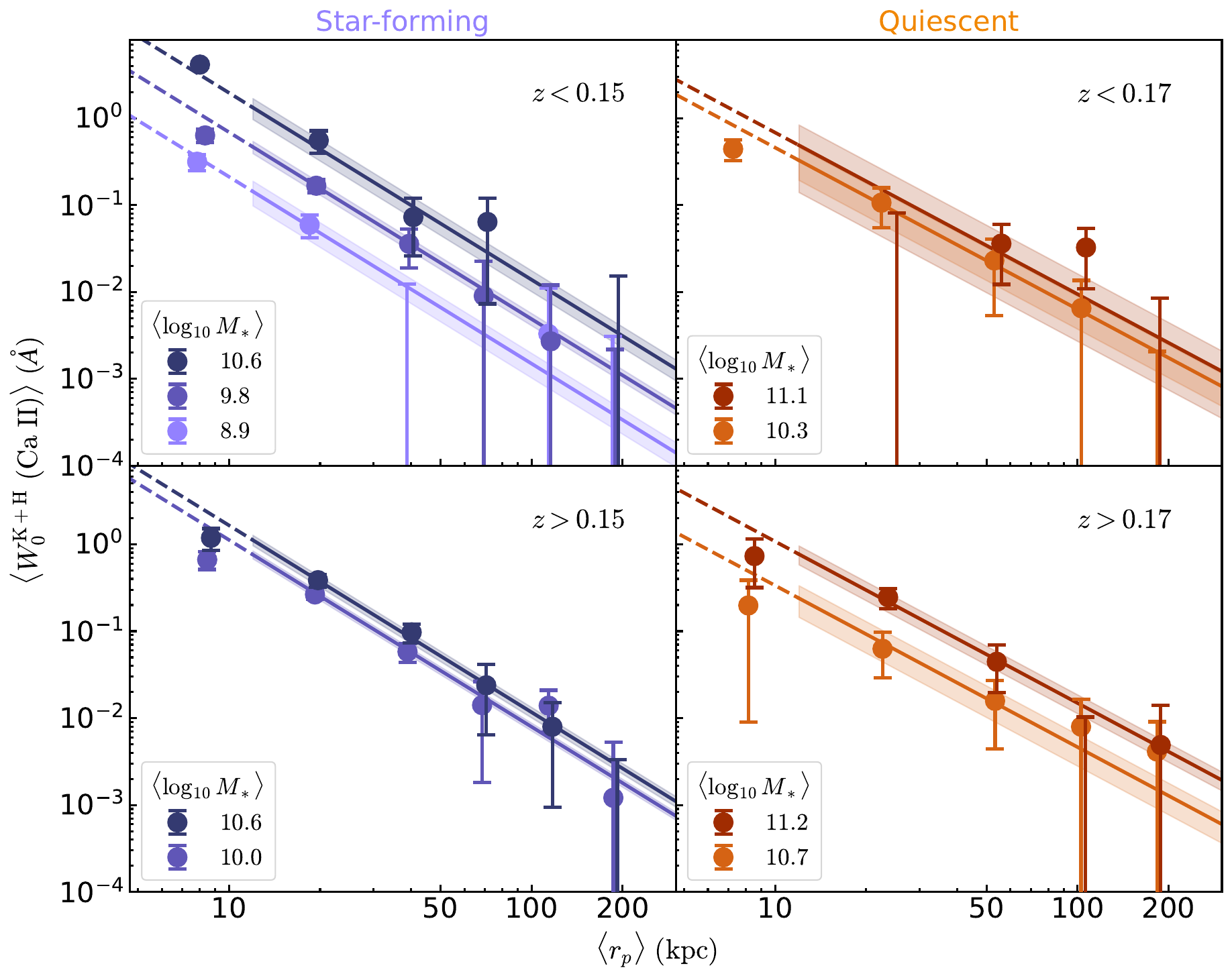}
\caption{Dependence of the \caii K and H equivalent width on stellar mass as a function of impact parameters, separating the samples into two redshift bins: $z=0.15$ for star-forming galaxies (left), and $z=0.17$ for quiescent galaxies (right). The median stellar masses are listed in the lower left. The solid lines follow the exponents fitted in Figure~\ref{fig:4mass}.} 
\label{fig:redshift_mass}
\end{figure*}

\begin{figure*}[htb!]
\centering
\includegraphics[width=\textwidth]{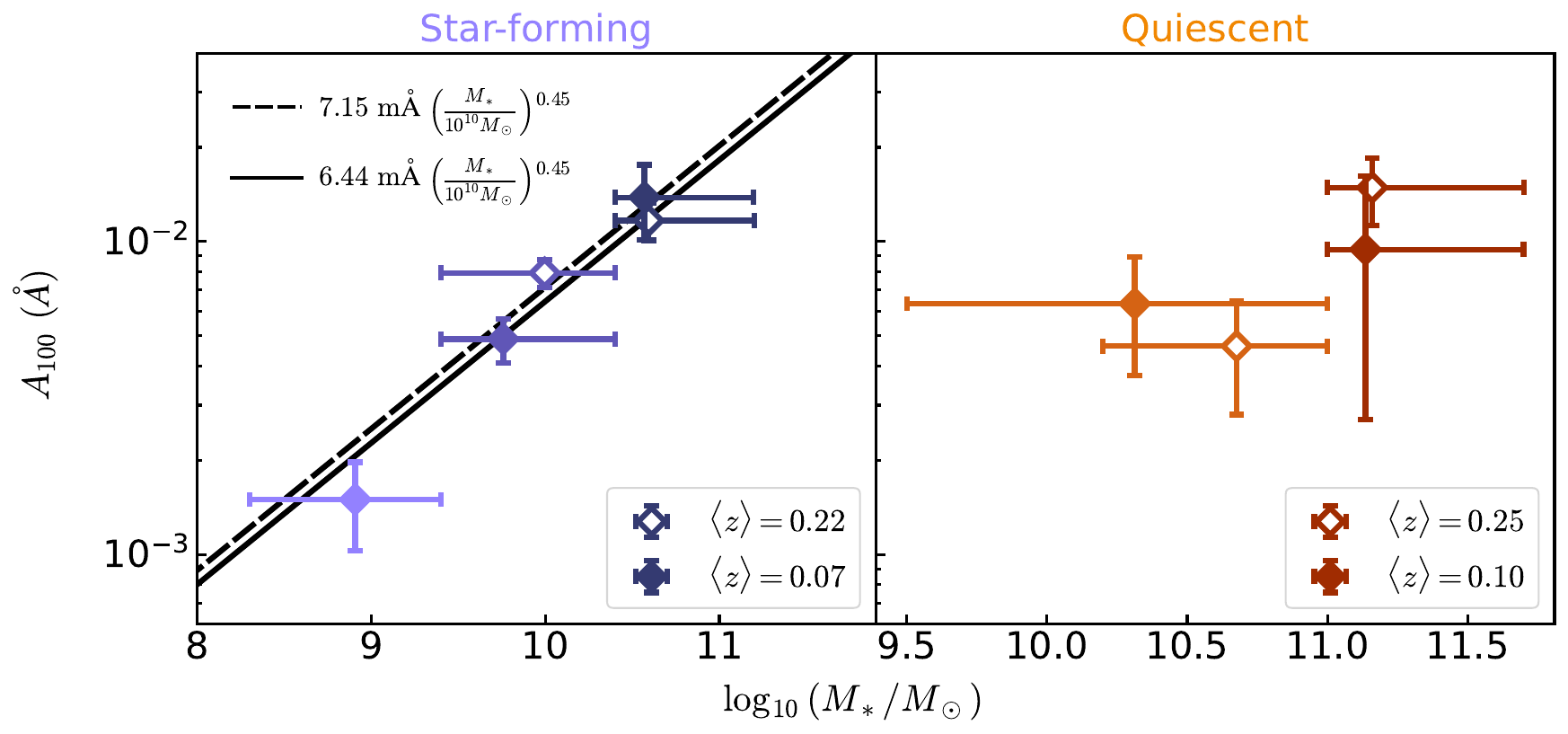}
\caption{Best-fit amplitudes as a function of median stellar masses for two redshift bins. The solid and open points represent low- and high-redshift samples, respectively. The solid and dashed lines in the left panel are the best-fit power laws for the low- and high-redshift samples.}
\label{fig:redshift_amplitude}
\end{figure*}

\begin{figure}[htb!]
\centering
\includegraphics[width=\columnwidth]{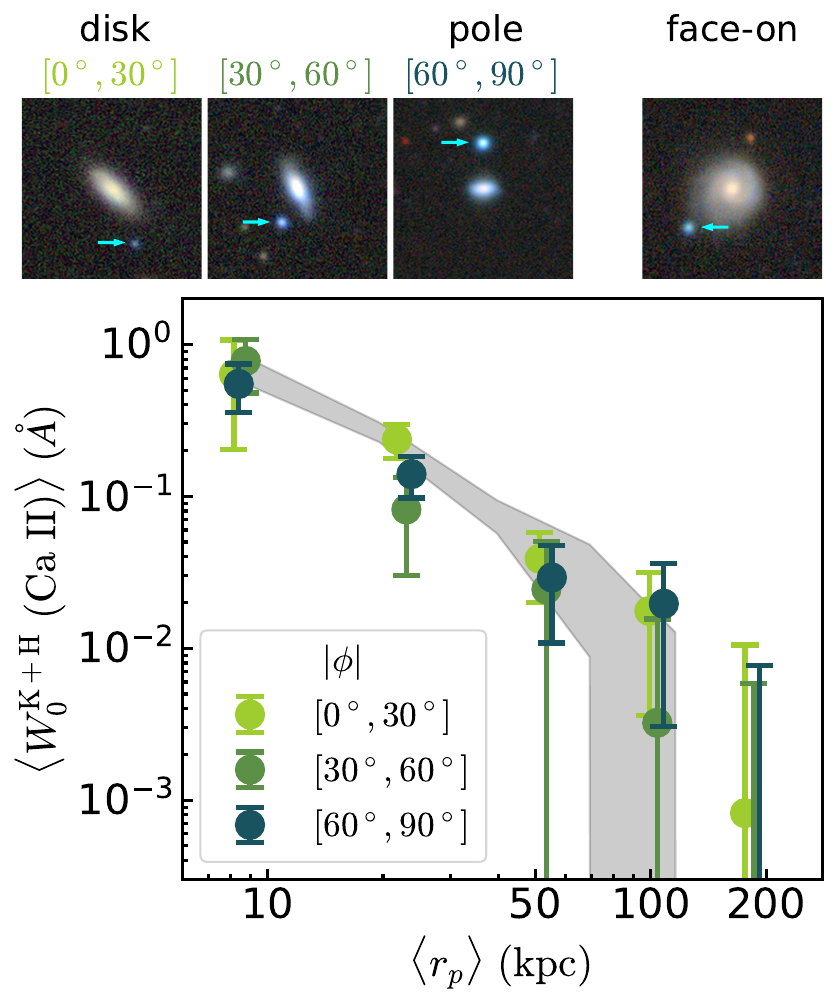}
\caption{Dependence of \caii on the azimuthal angles for galaxies with $\log_{10}(M_*/M_\odot)\ge 9.5$. Examples of the the galaxy--quasar pairs are shown at the top: from left to right are edge-on galaxies categorized into three azimuthal angles: disk, between disk and pole, pole; followed by face-on galaxies. The galaxies are located at the center, while the quasars are indicated by the cyan arrows. In the bottom panel, the gray band shows the  measurements of the face-on samples.}
\label{fig:azi}
\end{figure}

\subsection{Redshift dependence} 
With the correlations between \caii absorption and galaxy properties explored, we now investigate the correlation between \caii absorption and redshift as a function of stellar mass. We construct separate samples of star-forming and quiescent galaxies using redshift cuts at $z=0.15$ and $z=0.17$, respectively. In addition, stellar mass ranges are chosen so that redshift evolution has minimal impact across different mass bins. For star-forming galaxies, we include those with stellar masses in the range $10^{8.3} - 10^{11.2} M_\odot$ at $z<0.15$, and $10^{9.4} - 10^{11.2} M_\odot$ at $z>0.15$. For quiescent galaxies, we select those with stellar masses between $10^{9.5}$ and $10^{11.7} M_\odot$  at $z<0.17$, and between $10^{10.2}$ and $10^{11.7} M_\odot$ at $z>0.17$.
The left and right panels of Figure~\ref{fig:redshift_mass} show the results for star-forming and quiescent galaxies, while the upper and lower panels show the results for the low- and high-redshift bins, respectively. We also obtain the best-fit amplitudes of power laws (Equation~\ref{eqn:powerlaw}) with fixed $\alpha$ parameters obtained previously for star-forming and quiescent galaxies. The best-fit amplitudes are shown in Figure~\ref{fig:redshift_amplitude}. The solid and open data points indicate the measurements at low and high redshifts, respectively. 

For star-forming galaxies, the best-fit amplitudes $A_{100}$ of the \caii absorption strengths increase with stellar masses for both redshift bins. In addition, with a fixed stellar mass, there is a tentative trend showing that the best-fit amplitudes $A_{100}$ at $z>0.15$ are slightly higher than those at $z<0.15$. We quantify this trend with
\begin{equation}
    A_{100} = C\left(\frac{M_{*}}{10^{10} \, M_{\odot}}\right)^{\beta}(1+z)^\delta,
      \label{eqn:redshift}
\end{equation}
and find $C=6.11\pm1.58\,\mathrm{m\AA}$, $\beta=0.45\pm0.10$, and $\delta=0.80\pm1.57$.
We note that despite the uncertainty, the redshift trend is similar to the redshift evolution observed in the cool gas around galaxies traced by \mgii absorption lines \citep[e.g.,][]{lan20, schroetter21, cherrey25}. For example, \cite{lan20} showed that the covering fraction of \mgii absorption lines with $W_{\lambda2796}>1\, \rm \AA$ around star-forming galaxies evolve with redshift as $(1+z)^{2.2\pm0.4}$. On the other hand, we do not observe any redshift dependence for \caii around quiescent galaxies, which is inconsistent with the results of \cite{lan20} for quiescent galaxies. This might indicate that the cool gas around passive and star-forming galaxies evolve differently. However, further confirmation of such a scenario requires higher S/N measurements of \caii absorption. 

In \cite{lan20}, the author proposed that the redshift evolution of the \mgii covering fraction is connected with the redshift evolution of SFR of galaxies \citep[e.g.,][]{whitaker12, scoville17, koprowski24} given that they have similar redshift evolution. 
For our \caii measurements, we also estimate the SFR of star-forming galaxies at two redshift bins with stellar mass $>10^{10} M_\odot$ and find that the ratio between the median SFR, $\mathrm{SFR}(z>0.15)/\mathrm{SFR}(z<0.15)$, is $\sim 1.6$. Together with our results indicating a SFR and absorption strength correlation with $\propto \rm SFR^{0.31}$ (Figure~\ref{fig:sfr}), the absorption strength difference ratio is expected to be $\sim 1.6^{0.31} \sim 1.16$, which is consistent with the redshift trend observed in Figure~\ref{fig:redshift_amplitude}. This consistency suggests that the redshift evolution of \caii absorption around star-forming galaxies might be closely related to the star formation activities of galaxies.

\subsection{Azimuthal angle dependence} \label{sec:azi}
One can further investigate possible mechanisms contributing to gas observed in the CGM by measuring gas absorption around star-forming galaxies as a function of azimuthal angle. This is motivated by the fact that galaxy simulations predict that galactic outflows driven by supernova and black hole feedback preferentially escape galaxies along the minor axis \citep[e.g.,][]{nelson19}. To do so, we examine the azimuthal dependence of our sample by choosing star-forming galaxies with $\log_{10}\left({M_*/M_\odot}\right) \ge 9.5$ and with inclination angle larger than $60 ^\circ$. 
The inclination angles of galaxies and azimuthal angles between galaxy--quasar pairs are calculated based on the shape information measured via \textsc{Tractor}\footnote{\url{https://www.legacysurvey.org/dr9/catalogs/}} \citep{lang16} applied to the DESI Legacy Imaging Surveys \citep{dey19}. In general, the \textsc{Tractor} performs model-based photometry by simultaneously fitting sources across multiple bands and classifies them as point or extended sources with the point-spread function taken into account. Extended sources are further modeled using exponential, de Vaucouleurs, or composite profiles to best represent their light distribution. To evaluate the precision of the shape parameter measurements, we cross-match approximately 1500 DESI galaxies with inclination angles larger than $60 ^\circ$ to the Euclid Q1 data set \citep{euclid25}, and compare the position angles used in this work and those from Euclid Q1. The result shows that approximately 95\% of galaxies exhibit differences smaller than $10^\circ$.
We then divide the selected sample into three azimuthal angle bins: $|\phi| \in [0^\circ, 30^\circ]$, $[30^\circ,  60^\circ]$ and $[60^\circ, 90^\circ]$, which correspond to the disk, between disk and pole, and the pole of the foreground galaxies. Galaxy image examples can be found in the upper panel of Figure~\ref{fig:azi}, where the foreground galaxies are located at the centers, while the background quasars are pointed out by cyan arrows.

The results are shown in the lower panel of Figure~\ref{fig:azi} with colors indicating measurements for different azimuthal angles and the gray band for the measurements of the face-on galaxies. There is no significant correlation between \caii absorption and the azimuthal angle. This result is inconsistent with previous results of the cool CGM obtained at higher redshifts, showing preferentially stronger \mgii absorption along the minor axis than along the major axis within 50 kpc \citep[e.g.,][]{bordoloi11, bouche12, kacprzak12, lan14, lan18, zabl19}. In addition, our results are inconsistent with \caii measurements from \cite{zhu13}. The authors reported that the absorption strength of \caii is stronger along the minor axis than the major one. However, we note that the conclusion in \cite{zhu13} is driven by a single measurement at $r_p \sim 15\ \mathrm{kpc}$, a trend that is not observed in our measurements with more pairs. We will further discuss the implications of our azimuthal angle measurements in Section~\ref{subsec:dis_azu}.

\subsection{Gas distribution with respect to the size of halos} \label{sec:vir}

Given that the same impact parameter in the physical space corresponds to different regions in the halos of galaxies with different masses, we measure the \caii absorption as a function of impact parameter normalized by the virial radius, $r_p/r_\mathrm{vir}$, of the dark matter halos of the central galaxies. To estimate the virial radius, we follow \cite{chang25} and derive halo masses using Bayesian inference. Briefly, the halo masses are estimated from stellar masses combining information of a prior and a likelihood function, which are constructed using the halo mass function from \cite{tinker10} and the stellar-to-halo mass relation from \cite{moster13}, respectively. For further details, we refer the readers to \cite{chang25}. Using the estimated halo masses, we determine the virial radius with the analytical formula from \cite{bryan98}.

\begin{figure*}[ht!]
\centering
\includegraphics[width=\textwidth]{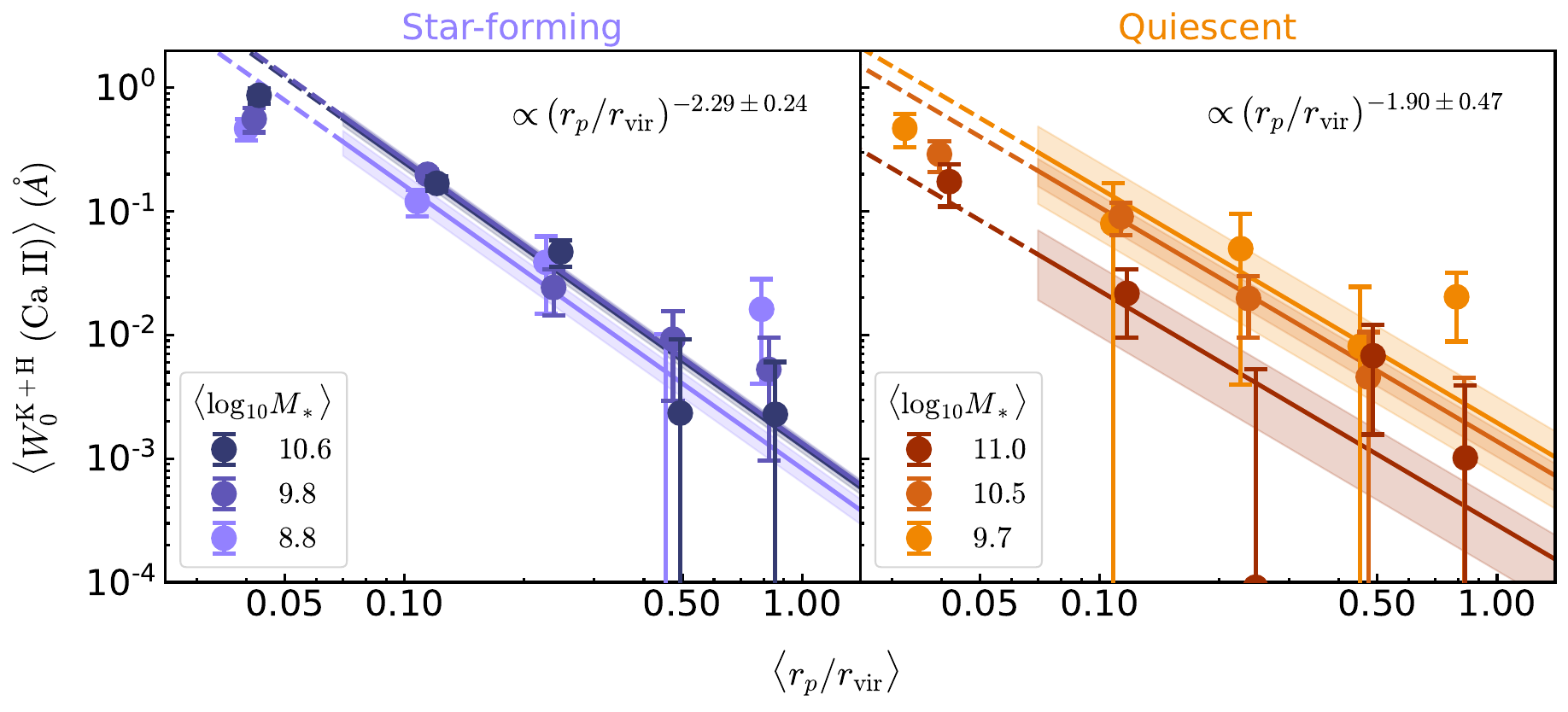}
\caption{Dependence of the \caii K and H equivalent width on stellar masses as a function of normalized distances. Left: star-forming galaxies. Right: quiescent galaxies. The median stellar masses are listed in the lower left.}
\label{fig:vir}
\end{figure*}

The results are shown in Figure~\ref{fig:vir}. The \caii measurements around star-forming and quiescent galaxies are shown in the left and right panels, respectively, fitted with
\begin{equation}
    \langle W_0 \rangle\left(\frac{r_p}{r_\mathrm{vir}}\right) = A_{0.5\rm vir}\left(\frac{r_p}{0.5r_\mathrm{vir}}\right)^\eta. \label{eqn:powerlaw_vir}
\end{equation}
We find that after the normalization, \caii absorption profiles around star-forming galaxies with different stellar masses become consistent with each other, indicating that the cool gas distributes similarly in the halos, and the behavior of the cool CGM is closely tied to the halo masses of the host galaxies. On the other hand, for quiescent galaxies, there is a different trend, showing that the absorption strengths decrease with stellar masses, especially in the inner regions of the halos $<0.3 \, r_\mathrm{vir}$. In addition, again, for galaxies with $>10^{10}\, M_{\odot}$, the \caii absorption around star-forming galaxies is approximately 1.5--2 times stronger than that around quiescent galaxies as observed in the physical space. These results demonstrate that while cool gas traced by \caii exists in both types of galaxies, they behave differently around different types of galaxies. 

Our \caii results are consistent with results obtained from \mgii absorption lines, showing that the \mgii absorption profiles around star-forming galaxies with different masses are similar when normalized by the distance $r_p/r_\mathrm{vir}$, while the absorption decreases with the mass of galaxies for quiescent ones \citep[e.g.,][]{churchill13, churchill13b, lan20, anand21}. 

\subsection{\caii mass in the halos} \label{sec:CaII_mass}
\begin{figure}[ht!]
\centering
\includegraphics[width=\columnwidth]{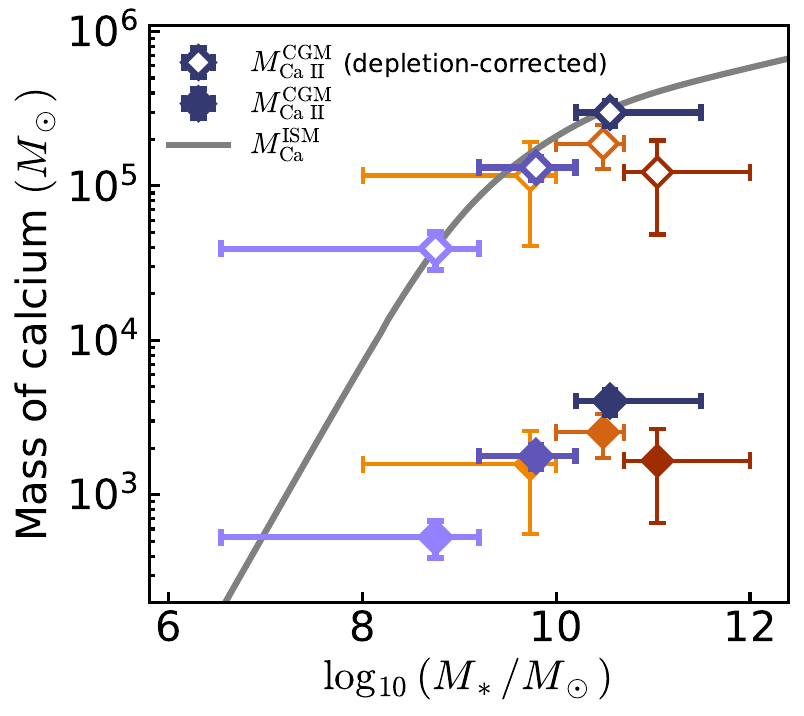}
\caption{\caii mass in CGM. The solid data points are estimated \caii mass around star-forming galaxies (blue) and quiescent galaxies (red) and the open data points are values corrected for dust depletion. The gray line is the Ca mass in ISM inferred from the mass--metallicity relation and atomic mass sequence.}
\label{fig:CaII_mass}
\end{figure}

We now estimate the \caii mass in the halos. To this end, we first calculate the \caii column density based on the \caii K line rest equivalent width as $\langle W_{0}^\mathrm{K}\rangle=\frac{2}{3}\times \langle W_{0}^\mathrm{K+H}\rangle$ 
using the following equation adopted from \citep{draine11}
\begin{equation}
    N_\mathrm{Ca\,II} = 1.13\times 10^{12}\ \mathrm{cm^{-1}}\ \frac{W_0^\mathrm{K}}{f\lambda^2},
\end{equation}
where $f$ is the oscillator strength of \caii K being 0.648, and $W_0^\mathrm{K}$ and $\lambda$ are in units of $\rm\AA$. This relation works for unsaturated, optically thin absorption lines and provides a lower limit of column density when the lines are saturated. Our measurements are in the unsaturated region with the strongest $\langle W_{0}^\mathrm{K}\rangle$ being $ 0.2 \, \rm \AA$ at $r_{p}>10\,\mathrm{kpc}$ \citep[e.g.,][]{zhu13}. We then adopt the best-fit profiles as shown in Figure~\ref{fig:vir}, and integrate the \caii mass from $0.1 r_\mathrm{vir}$ to $1 r_\mathrm{vir}$: 
\begin{eqnarray}
    \langle M_\mathrm{Ca\,II} \rangle = 2\pi m_\mathrm{Ca} \int_{0.1r_\mathrm{vir}}^{r_\mathrm{vir}} N_\mathrm{Ca II}\left(r_p\right) r_p \mathrm{d}r_p,
    \label{eqn:caiimass}
\end{eqnarray}
where $m_\mathrm{Ca}$ is the atomic mass of Ca. The results are shown in Figure~\ref{fig:CaII_mass}. We emphasize that given that \caii only contains a fraction of Ca in the ionization state, the estimated \caii mass represents the lower limit of the total Ca mass. 

In addition, we estimate the total Ca mass in the ISM of star-forming galaxies by combining the stellar mass--metallicity relation and atomic gas sequence from \cite{scholte24} with the solar Ca-to-O abundance ratio of 0.004 \citep{solarabund}. We also apply a dust depletion correction to the metallicity using the relation from \cite{peimbert10}. The estimated total Ca mass in the ISM of star-forming galaxies is shown by the gray line. For the CGM, we also account for dust depletion in our \ion{Ca}{2} measurements of the halos by applying a depletion factor. Using \textsc{Cloudy} simulations \citep{cloudy23}, we obtain an estimated depletion factor of $\sim 74$ by comparing the intrinsic \ion{Ca}{2}-to-\ion{Zn}{2} ratio with the observed ratio in the gas traced by \mgii from \cite{lan17} assuming that \caii and \mgii mostly originate from the same gas in the CGM. This assumption is motivated by the fact that the combination of the \mgii covering fraction \citep{lan14} at $z\sim0.4$ and the typical \caii absorption strength in \mgii absorbers \citep{lan17} yields similar \caii absorption profiles around galaxies as observed in this work. A detailed description of the simulations is provided in Appendix \ref{sec:cloudy}. After the correction, the amount of the \caii mass in the halos is comparable to the amount of Ca mass in the ISM. Moreover, the relationship between the \caii mass and stellar mass of star-forming galaxies follows the relationship of the ISM metal mass and stellar mass of star-forming galaxies. This result demonstrates that the metal enrichment of the ISM and CGM is closely linked together, reflecting the cumulative effect of feedback mechanisms. 

\section{Discussion} \label{sec:discussion}

\subsection{Comparisons with previous results}

By combining DESI spectra of background quasars, we have measured the \caii absorption strengths around galaxies as a function of stellar mass, galaxy types, and redshift, and revealed correlations between \caii absorption strengths and galaxy properties. In the following, we compare our results with similar measurements from previous studies and discuss the implications. 

\textbf{Cool CGM traced by \caii at $z<0.2$:}
Our observed trends are consistent with the trends reported in the previous \caii results at $z\sim0.1$ using the SDSS data set by \citet{zhu13}. For example, \citet{zhu13} separated galaxies into two mass bins and found that galaxies with $10^{10.5}$ and $10^{9.8}\, M_{\odot}$ have best-fit $A_{100}$ parameter values of $15$ and $7 \, \rm m\AA$ at 100 kpc. This yields $\beta \sim \rm 0.47$, which is consistent with our best-fit $\beta$ parameter value. We note that while in their analysis, only a mass cut is applied, we expect that the trend in their measurements is mostly driven by signals from a star-forming population that has a larger pair number. In addition, they reported that \caii around star-forming galaxies is approximately 2 times stronger than \caii around quiescent galaxies, a trend that we have found in DESI data. 
We note that limited by the SDSS data set, \citet{zhu13} explored either the dependences between \caii absorption and mass or SFR, while the DESI data set enables us to perform the exploration as a function of mass and SFR independently across a wider parameter space than previously explored. We also note that the $\alpha$ parameter of our best-fit profiles is $\sim-2$ for both types of galaxies, while \citet{zhu13} obtained $\sim-1.4$. The difference is mostly due to the fact that we exclude the $<10\ \, \rm kpc$ measurements during the fitting, while \citet{zhu13} included the measurements in their fitting. In Appendix \ref{sec:sdss}, we directly compare DESI measurements with SDSS measurements from \citet{zhu13} and show that the measurements from the two data sets are consistent. 

The only inconsistency between our results and the results in \citet{zhu13} is the azimuthal angle dependence. Given that the trend observed in \citet{zhu13} is driven by a single measurement in the innermost bin, it is possible that such a trend is affected by small statistics limited by the SDSS data. The larger DESI sample is expected to be less sensitive to the effect. Our measurements are also consistent with the \caii measurements within 10 kpc of galaxies obtained by \citet{rubin22}\footnote{We multiply the \caii K measurements in \cite{rubin22} by a factor of 1.69, which is the weighted mean of the ratio of H+K to K lines from the innermost bins of our measurements.} as shown in Figure~\ref{fig:4mass}, as well as the \caii measurements in our Milky Way ISM with typical $W_{0}^{\rm H+K}\sim 0.8-1 \rm \, \AA$ from \citet{murga15}\footnote{Given that the Milky Way sightlines intercept approximately half of the ISM, we multiply the \caii measurements by two.}.  

\textbf{Cool CGM traced by \mgii at $z\gtrsim 0.4$:} 
Explorations of the galaxy--cool CGM connections have been conducted with \mgii absorption lines as the tracers of the cool CGM \citep[e.g.,][]{bordoloi11, kacprzak12, rubin14, nielsen16, dutta20, huang21}, and have revealed similar correlations as found in this work:
\begin{itemize}
    \item \textbf{Mass dependence}: Studies have shown that the covering fraction and the absorption profiles of the cool CGM traced by \mgii absorbers increase with stellar mass (and halo mass) of systems. For example, using $\sim 100$ galaxy--quasar pairs, \citet{chen10} showed that the characteristic radius of gas traced by \mgii scales with galaxy $B$-band luminosities as $\propto L_{B}^{0.35}$. With $\sim200$ galaxy--quasar pairs obtained with MUSE observations, \citet{dutta20} also concluded that the stellar mass is the primary factor affecting \mgii absorption around galaxies. 
    Similarly, via a statistical cross-correlation analysis, \citet{lan20} demonstrated that the covering fraction of \mgii absorbers measured in physical space correlates with the galaxy mass with $\propto M_{*}^{0.5}$ for star-forming galaxies. 
    Finally, based on cool gas measurements around galaxies and quasars living in halos with different masses, \citet{prochaska14} also argued that the halo mass is the primary factor affecting the cool gas distribution, a conclusion that is also reached by \citet{churchill13}. Together with the \caii measurements, this mass dependence of the cool CGM has been observed across a wide range of redshifts from $z\sim0$ up to $z\sim2$, covering at least 10 Gyr cosmic time. This provides hints on the underlying mechanisms that are expected to operate from $z\sim2$ and persist through cosmic time. 

    \item \textbf{Galaxy type dependence:} In addition to the mass dependence, the correlation between galaxy types and the cool CGM, especially in the inner regions, has been reported. Several studies show that the average \mgii absorption strength in the inner CGM around star-forming galaxies is stronger than that around quiescent galaxies \citep[e.g.,][]{bordoloi11, lan14, rubin18, lan20, anand21, huang21}. A difference of a factor of $\sim4$ is observed within 50 kpc, and the value is larger with smaller impact parameters. We note that such a trend is found in \caii measurements as shown in Section~\ref{sec:results}. However, the differences in \caii absorption between star-forming and quiescent galaxies are at most a factor of 2 at $r_{p}\sim20$ kpc, which is smaller than the results from \mgii absorption at higher redshifts. This might suggest that the dependence of the cool CGM on galaxy types becomes weaker at lower redshifts, possibly associated with the declines of the SFR of galaxies and the contribution of galactic outflows. We discuss this implication in the following subsection.  
\end{itemize}

\subsection{Possible redshift evolution of the azimuthal angle dependence} \label{subsec:dis_azu}
Blueshifted absorption lines, the observed signature of galactic outflows, have been detected ubiquitously in the spectra of star-forming galaxies, via the so-called down-the-barrel observations \citep[e.g.,][]{weiner09,rubin14,zhu15}. With the typical observed velocity ($\sim 300\,\mathrm{km\,s^{-1}}$; e.g., \citealt{rubin14}), gas associated with outflows is expected to travel into the halos and possibly reaches to 50 kpc within a few hundred megayears \citep[e.g.,][]{lan19}. The azimuthal angle dependence of gas distribution has been considered as one of the possible imprints of galactic outflows in the CGM. Such a distribution has been observed in the cool gas traced by \mgii absorption lines. For instance, \citet{bordoloi11} measured the \mgii absorption along the minor and major axes of star-forming galaxies at $z\sim0.7$ with Hubble Space Telescope Advanced Camera for Surveys images and found that within 50 kpc, the \mgii absorption is stronger along the minor axis than the major axis. The rest equivalent width of \mgii is $\sim 3$ times higher along the minor axis than the major axis at $\sim25$ kpc. A similar trend and the ratio between \mgii rest equivalent width along the minor and major axes are also found in \citet{lan18} at $z\sim 0.8$. 
However, we note that at $z<0.4$ whether or not the azimuthal angle dependence is as strong as the signals observed at $z>0.4$ is still under debate. For example, at $z\sim0.2$, \citet{bouche12} and \citet{kacprzak12} reported a bimodal azimuthal dependence of \mgii absorption. \citet{martin18} also reported strong \mgii absorption found along both the minor and major axes of galaxies, showing a smaller ratio of the \mgii rest equivalent width along the minor to major axes than the ratio observed at $z>0.4$ \citep[e.g.,][]{bordoloi11, lan18}.
On the other hand, at similar redshift, \citet{huang21} reported no significant azimuthal dependence on the covering fraction and absorption strength of \mgii absorbers. Apart from \mgii studies, \citet{pointon19} also reported a lack of azimuthal trends for metallicity distributions traced by Si around galaxies at $z\sim0.27$. At lower redshifts, \citet{borthakur15} found no azimuthal dependence of Ly$\alpha$ absorption at $z\sim0.037$, although their sample probes relatively large impact parameters. Our results align with these latter studies, showing no azimuthal angle dependence for gas traced by \ion{Ca}{2}. The discrepancy might result from the sample differences and the gas tracers used in different studies, yet the details remain to be clarified, calling for future research to complete the picture.

Despite of the abovementioned debate, the current results suggest that the degrees of azimuthal angle dependence vary with different redshifts, being weak or nondetectable at the low-redshift CGM while observed mostly at $z>0.4$ \citep{bordoloi11,lan18}. We propose that this trend may result from the redshift evolution of galactic outflows driven by star formation activities.
For star-forming galaxies with similar mass, the average SFR at $z\sim0.8$ is $\sim 3$ times higher than that at $z\sim0.2$ \citep[e.g.,][]{moustakas13}. Therefore, three times more gas mass associated with outflows is expected to contribute to the CGM preferentially allocated along the minor axis. Moreover, the velocities of galactic outflows evolve with redshifts. With a fixed SFR, the maximum outflow velocity at $z\sim 1$ is approximately 1.5--2 times faster than the outflow velocity at $z\sim 0$ \citep[e.g.,][]{sugahara19}. Taking these two factors together, at higher redshifts, galactic outflows are expected to distribute more gas mass into the CGM at larger distances. In other words, the absence of azimuthal angle dependence observed in \caii at low redshifts might indicate that gas associated with galactic outflows does not contribute significantly in the inner CGM. This lack of galactic outflow contribution can also explain why \caii absorption in the inner CGM around star-forming and quiescent galaxies at $z\sim0.2$ does not have such a large difference observed at $z\sim0.7$ with \mgii absorption. 

This suppression of galactic outflows toward low redshifts is aligned with the results from the Feedback In Realistic Environments (FIRE) simulations \citep{fire}. \citet{muratov15} and \citet{angles17} found that high-redshift galaxies drive strong galactic outflows produced by bursty star formation, while at low redshifts, massive galaxies shift into steady star formation that does not drive strong outflows. \citet{stern21} proposed that this transition of the star formation mode is linked to the virialization of the inner CGM, which occurs at lower redshifts, and after the virialization of the outer CGM. Before the virialization of the inner CGM, the inner CGM is dominated by clumpy cool gas, making galactic outflows relatively easy to travel through. After the virialization of the inner CGM, the hot gas around galaxies confines the galactic outflows. This transition of the inner CGM virialization for $L^*$ galaxies occurs at $z\sim0.8-1$ in the FIRE simulations, which coincides with the epoch separating the behaviors of cool gas around star-forming galaxies. While further testing and validation of this outside-in CGM virialization scenario requires direct comparisons between the predicted and observed properties of cool gas tracers, such as \mgii and \ion{Ca}{2}, around galaxies across time and mass, the current observed properties of the cool CGM are consistent with the results of the FIRE simulations qualitatively. 

\subsection{Impact of AGNs on the properties of the CGM}
Our results show that there is no significant difference in the \caii absorption between AGNs and star-forming galaxies. This indicates that current AGN activities do not have strong impact on the properties of the cool CGM. This result is consistent with previous measurements of gas around galaxies with AGN activities. For example, \citet{berg18} selected 19 AGNs via BPT diagnosis and measured the CGM properties with COS UV spectroscopy, finding that the absorption-line properties of AGNs are consistent with those of the control sample within 160 kpc. Similarly, \citet{kacprzak15} estimated the covering fraction of \mgii absorption around 14 AGNs, and obtained a value slightly lower than that of field galaxies, though within the uncertainties. For luminous AGNs, i.e., quasars, the cool gas distribution has also been measured via quasar--quasar pairs. Previous results show that the properties of cool gas around quasars are similar to those around galaxies with similar stellar mass, suggesting that the halo mass is the primary factor driving the properties of the cool gas \citep[e.g.,][]{farina14, prochaska14}. On the other hand, \citet{johnson15} reported a correlation between the covering fraction of \mgii absorption and quasar luminosity and argued that quasar feedback can be one of the mechanisms driving the correlation. While validating such a physical connection requires further investigations, it suggests that  energy from powerful/luminous quasars may be needed to sufficiently leave imprints on the properties of the cool CGM. Finally, considering radio-mode feedback, \citet{chang24} measured the properties of the cool gas traced by \mgii around radio galaxies and the control sample and found no detectable difference between the properties of the gas around the two populations, illustrating that recent radio AGN activities do not impact the properties of the cool gas strongly. 
These measurements again support the picture that the properties of galaxies, especially the stellar/halo mass and SFR, are the dominant factors correlating with the properties of the cool gas. 

\section{Conclusions} \label{sec:conclusion}

Utilizing the DESI Y1 galaxy and quasar spectroscopic data sets, we constructed high S/N composite spectra of quasars in the background of galaxies from which we detected and measured the absorption strengths of \ion{Ca}{2}, tracers of cool gas, around galaxies. 
We explored the relationships between the properties of the cool CGM and the properties of galaxies, including stellar mass, SFR, redshift, and AGN activities, and compared our measurements at $z<0.4$ with similar measurements obtained at higher redshifts. Our results are summarized in the following:

\begin{enumerate}
    \item We find that the \caii rest equivalent widths decrease with impact parameters, consistent with the cool CGM measurements from previous studies.
    
    \item For star-forming galaxies, the absorption strengths correlate with the stellar mass of galaxies from $10^{8} \, M_{\odot}$ to $10^{11}\, M_{\odot} $ with $\propto M_{*}^{0.5}$, while such a trend is weaker for quiescent galaxies. With a fixed stellar mass, we also find a positive correlation between absorption strengths and SFR for star-forming galaxies, following $\propto\mathrm{SFR}^{0.3}$. In addition, for $>10^{10} \, M_{\odot}$ galaxies, the absorption strengths are stronger around star-forming galaxies than those around quiescent galaxies within $30\,\mathrm{kpc}$. The correlations between \caii and galaxy properties are detected at $z\sim0.1$ and $z\sim0.3$. 
    
    \item We find that there is no azimuthal angle dependence of absorption strength around star-forming galaxies, which is inconsistent with previous measurements at higher redshifts. We argue that this might be due to the redshift evolution of galactic outflows which is expected to carry less mass with slower velocity at lower redshifts.

    \item We measure \caii absorption around AGNs, selected with the BPT diagnosis, and find that with similar galaxy masses, \caii absorption strengths around AGNs and star-forming galaxies are consistent. This result suggests that current AGN activities do not have strong impacts on the properties of the cool CGM. 

    \item After normalizing the impact parameter by the virial radius, the absorption profiles around galaxies with different stellar masses from $10^{8} \, M_{\odot}$ to $10^{11} \, M_{\odot}$ become consistent with each other, indicating that the cool CGM distributes similarly in the halos and is closely linked to halo masses. For quiescent galaxies with masses $>10^{11} \, M_{\odot}$, the \caii absorption in the inner CGM appears to be weaker. This halo mass trend is also observed at higher redshifts, suggesting that the underlying mechanism(s) regulating such a property persists across cosmic time. 
    
    \item We estimate the mass of \caii as a function of stellar mass, and find that after the dust depletion correction, the amount of \caii in the CGM is comparable with the Ca mass in the ISM of galaxies. Moreover, the cool CGM metal mass and stellar mass relation of star-forming galaxies follows the ISM metal mass and stellar mass relation. 
\end{enumerate}
Our results provide a comprehensive characterization of galaxy--cool CGM connections below $z<0.4$ down to $10^{8}\, M_{\odot}$ galaxies, offering novel constraints on the models of galaxy evolution. Together with similar measurements at higher redshifts, the relationships between the properties of galaxies and the cool CGM have been mapped across $\sim10$ Gyr from redshift 0 to redshift $\sim2$. Understanding the physical mechanisms behind these relationships over cosmic time will be essential for advancing the knowledge of galaxy formation and evolution.

The analysis adopted in this work can be directly applied to upcoming data sets. For example, at the local Universe, the complete 5 yr DESI data set is expected to further enhance the S/N of the similar measurements by at least a factor of two. In addition, one can combine the images provided by Euclid \citep{euclid} to obtain more precise measurements of gas distribution as a function of azimuthal angle, a key constraint on how gas flows around galaxies. Similar measurements will also be obtained at even higher redshifts. The Prime Focus Spectrograph survey \citep{pfs} and the DESI-II \citep{desi-ii}, for example, will observe millions of galaxies at $z>2$, enabling explorations of galaxy--CGM connections with large data sets at the peak of the star formation density of the Universe \citep[e.g.,][]{madau14}. Similar to the properties of galaxies, such as stellar mass functions and luminosity functions at different redshifts \citep[e.g.,][]{kelvin14, driver22, weaver23}, precise measurements of gas around galaxies at different redshifts will serve as fundamental constraints, leading to a more comprehensive understanding of the cosmic baryon cycle. 

\section*{Data Availability}
All data points shown in the figures are available at Zenodo DOI:\dataset[10.5281/zenodo.17083730]{https://doi.org/10.5281/zenodo.17083730}.

\begin{acknowledgments}
We thank the anonymous referee for the constructive report.
We also want to thank Siwei Zou and Abhijeet Anand for their constructive comments that greatly helped improve the paper during the DESI Collaboration Wide Review. 
Y.V.N., T.W.L., and Y.L.C. acknowledge support from the National Science and Technology Council (MOST 111-2112-M-002-015-MY3, NSTC 113-2112-M-002-028-MY3), the Yushan Fellow Program by the Ministry of Education (MOE) (NTU-110VV007, NTU-111V1007-2, NTU-112V1007-3, NTU-113V1007-4) and National Taiwan University research grant (NTU-CC-111L894806, NTU-CC-112L893606, NTU-CC-113L891806, NTU-111L7318, NTU-112L7302).

This material is based upon work supported by the U.S. Department of Energy (DOE), Office of Science, Office of High-Energy Physics, under Contract No. DE–AC02–05CH11231, and by the National Energy Research Scientific Computing Center, a DOE Office of Science User Facility under the same contract. Additional support for DESI was provided by the U.S. National Science Foundation (NSF), Division of Astronomical Sciences under Contract No. AST-0950945 to the NSF’s National Optical-Infrared Astronomy Research Laboratory; the Science and Technology Facilities Council of the United Kingdom; the Gordon and Betty Moore Foundation; the Heising-Simons Foundation; the French Alternative Energies and Atomic Energy Commission (CEA); the National Council of Humanities, Science and Technology of Mexico (CONAHCYT); the Ministry of Science and Innovation of Spain (MICINN), and by the DESI Member Institutions: \url{https://www.desi.lbl.gov/collaborating-institutions}.

The DESI Legacy Imaging Surveys consist of three individual and complementary projects: the Dark Energy Camera Legacy Survey (DECaLS), the Beijing-Arizona Sky Survey (BASS), and the Mayall $z$-band Legacy Survey (MzLS). DECaLS, BASS and MzLS together include data obtained, respectively, at the Blanco telescope, Cerro Tololo Inter-American Observatory, NSF’s NOIRLab; the Bok telescope, Steward Observatory, University of Arizona; and the Mayall telescope, Kitt Peak National Observatory, NOIRLab. NOIRLab is operated by the Association of Universities for Research in Astronomy (AURA) under a cooperative agreement with the National Science Foundation. Pipeline processing and analyses of the data were supported by NOIRLab and the Lawrence Berkeley National Laboratory. Legacy Surveys also uses data products from the Near-Earth Object Wide-field Infrared Survey Explorer (NEOWISE), a project of the Jet Propulsion Laboratory/California Institute of Technology, funded by the National Aeronautics and Space Administration. Legacy Surveys were supported by the Director, Office of Science, Office of High Energy Physics of the U.S. Department of Energy; the National Energy Research Scientific Computing Center, a DOE Office of Science User Facility; the U.S. National Science Foundation, Division of Astronomical Sciences; the National Astronomical Observatories of China, the Chinese Academy of Sciences and the Chinese National Natural Science Foundation. LBNL is managed by the Regents of the University of California under contract to the U.S. Department of Energy. The complete acknowledgments can be found at \url{https://www.legacysurvey.org/}.

Any opinions, findings, and conclusions or recommendations expressed in this material are those of the author(s) and do not necessarily reflect the views of the U. S. National Science Foundation, the U. S. Department of Energy, or any of the listed funding agencies.

The authors are honored to be permitted to conduct scientific research on Iolkam Du’ag (Kitt Peak), a mountain with particular significance to the Tohono O’odham Nation.
\end{acknowledgments}

\appendix
\section{Cloudy simulation} \label{sec:cloudy}
\begin{figure}[ht!]
\centering
\includegraphics[width=\textwidth]{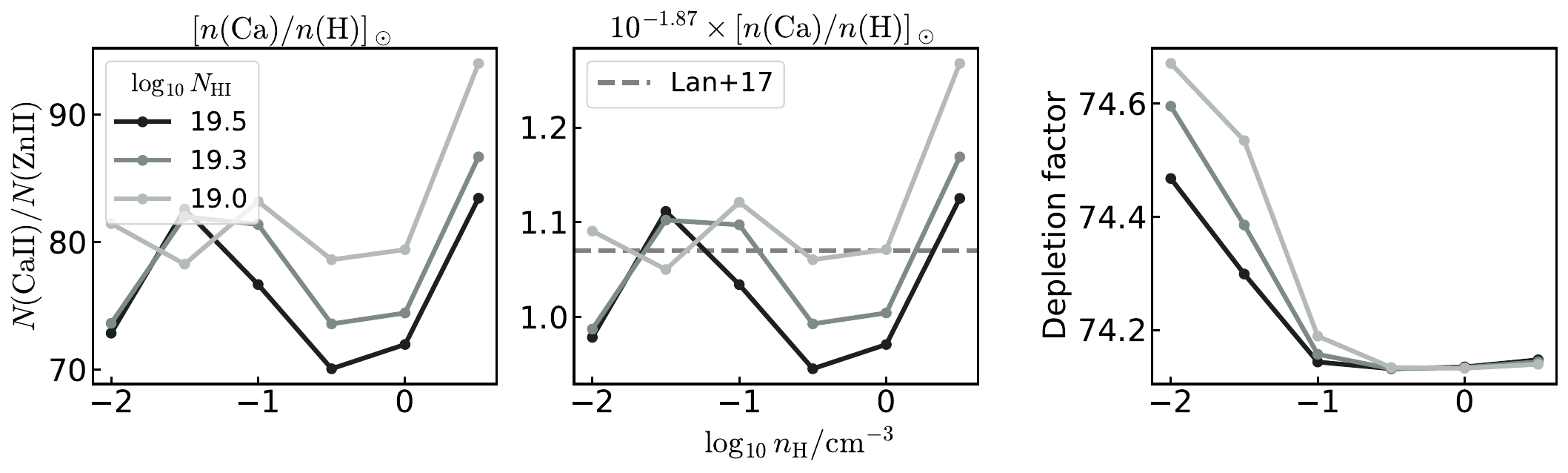}
\caption{\textsc{Cloudy} simulation results across a total hydrogen volume density range of $10^{-2}$--$10^{0.5}\,\mathrm{cm^{-3}}$ for $\log_{10}{(N_\mathrm{HI}/\mathrm{cm^{-2}})}=19.0,19.3,19.5$. The left panel shows the intrinsic \ion{Ca}{2}-to-\ion{Zn}{2} column density ratio, assuming an initial solar abundance. The middle panel shows the observed ratio from (\citealt{lan17}; dashed line) and simulated ratios by reducing the calcium abundance by a factor of $10^{-1.87}$. The right panel presents the depletion factor, obtained by dividing the intrinsic ratio from the left panel by the simulated depleted values from the middle panel.}
\label{fig:cloudy}
\end{figure}

We perform \textsc{Cloudy} simulations \citep{cloudy23} to evaluate the dust depletion factor in the CGM. The simulations are initialized with the \cite{haardt12} extragalactic radiation background at $z=0.2$, and run over a hydrogen volume density range of $10^{-2}$--$10^{0.5}\,\mathrm{cm^{-3}}$ until the neutral hydrogen column density reaches $10^{19}$, $10^{19.3}$, and $10^{19.5}\,\mathrm{cm^{-2}}$. In order to determine the depletion factor, we run two sets of simulations. First, we run a simulation by adopting the default solar composition in \textsc{Cloudy} \citep{grevesse98, prieto01,prieto02, holweger01} to obtain the intrinsic column density ratio of \caii and \ion{Zn}{2} (the left panel of Figure~\ref{fig:cloudy}). Next, we reduce the calcium abundance to simulate dust depletion, adjusting it to match the observed average \ion{Ca}{2}-to-\ion{Zn}{2} column density ratio of $\sim 1.07$ from (\citealt{lan17}; the middle panel of Figure~\ref{fig:cloudy}). The value is averaged from the \caii doublet values. By comparing the intrinsic ratio with the depleted simulations, we derive an average Ca depletion factor of $\sim 74$ and adopt this value to correct the depletion effect. 

\section{Comparison with SDSS} \label{sec:sdss}
\begin{figure}[hbt!]
\centering
\includegraphics[scale=0.6]{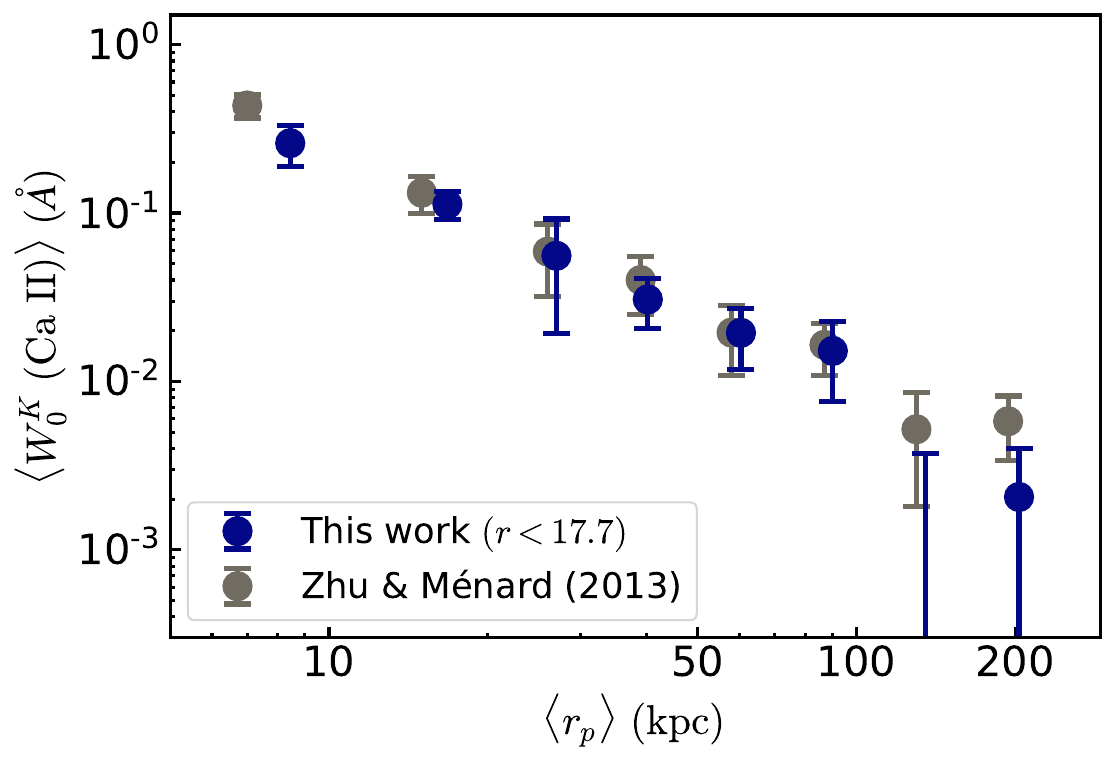}
\caption{Comparison of \caii absorption (K line) around $r < 17.7$ galaxies with SDSS and DESI data. The gray data points show the measurements from \citet{zhu13} and the blue data points show the measurements obtained in this work with DESI data.}
\label{fig:sdss}
\end{figure}

Here we compare our DESI measurements with the SDSS measurements from \citet{zhu13}. To match the SDSS galaxy population, we select DESI galaxies with $r < 17.7$ \citep{strauss02}. As shown in Figure~\ref{fig:sdss}, the DESI measurements are consistent with the SDSS measurements. We note that while DESI provides more galaxy--quasar pairs, the typical S/N of individual DESI quasar spectra is lower than the S/N of SDSS quasar spectra. Together, these two factors yield similar S/N values of the SDSS and DESI composite spectra. Nevertheless, the DESI data set enables us to probe the \caii absorption around galaxies with lower mass and higher redshifts. 

\bibliography{sample631}{}
\bibliographystyle{aasjournal}

\end{document}